\documentclass[10pt]{article}

\usepackage[margin=2cm]{geometry}
\usepackage{amsfonts}
\usepackage{graphicx} 
\usepackage{subfig}
\usepackage{cite}
\usepackage{caption}
\usepackage{amsmath}
\usepackage{amssymb}
\usepackage{color}
\usepackage[dvipsnames]{xcolor}
\usepackage{soul}
\usepackage[normalem]{ulem}
\usepackage{authblk}  
\usepackage[utf8]{inputenc}
\usepackage{epsfig}
\usepackage{float} 
\usepackage[toc,page]{appendix}
\usepackage{tikz}
\usetikzlibrary{patterns}
\usetikzlibrary{decorations.text}
\usetikzlibrary{decorations.markings}
\usetikzlibrary{decorations.pathmorphing}

\tikzset{snake it/.style={decorate, decoration=snake}}

\definecolor{EquationGreen}{HTML}{00aa00}
\definecolor{SCO}{HTML}{32CD32}
\definecolor{UCO}{HTML}{FFCC00}
\definecolor{Background}{HTML}{FAFAFA}
\definecolor{FontColor}{HTML}{23373B}
\definecolor{NoCO}{HTML}{8C1A8C}

\title{\textbf{Slowly Rotating Boson Stars}}

\author[1]{Jorge F. M. Delgado\footnote{jorgedelgado@icf.unam.mx}}

\author[1]{Juan Carlos Degollado\footnote{jcdegollado@icf.unam.mx}}

\affil[1]{\normalsize Instituto de Ciencias Físicas, Universidad Nacional Autónoma de México,
Apartado Postal 48-3, 62251, Cuernavaca, Morelos, México}

\author[2]{Luis E. Martínez
\footnote{enriquelmr@ciencias.unam.mx}}
\affil[2]{Department of Mathematics, Universitat Hamburg, Hamburg, 20146, Germany}

\author[3]{Marcelo Salgado\footnote{marcelo@nucleares.unam.mx}}
\affil[3]{Instituto de Ciencias Nucleares, Universidad Nacional Aut\'onoma de M\'exico, Circuito Exterior C.U., A.P. 70-543, M\'exico D.F. 04510, M\'exico.}

\begin{document}

\maketitle


\begin{abstract}
We present solutions to the Einstein-Klein Gordon system representing boson stars in the slow rotation approximation.  By considering slow rotation we are able to reduce the number of equations yielding a system of ordinary differential equations that is conveniently solved numerically without the need of expensive computational resources. We find sequences of  solutions and describe some of their physical properties such as, total mass, angular momentum and compactness. We also consider the dynamics of particles (geodesics) in the resulting spacetime. A detailed comparison with fully rotating boson stars (non-linear treatment) is performed by showing the region of validity of the slow-rotation approximation.
\end{abstract}

\tableofcontents

\newpage

\section{Introduction}
Boson stars (BSs) are globally regular, asymptotically flat, equilibrium solutions of the Einstein equations minimally coupled with a massive complex scalar field possibly with a self-interacting potential.
Spherically symmetric BSs with a massive but free scalar field
were first analyzed by Kaup \cite{Kaup:1968zz} and Ruffini and Bonazzola \cite{Ruffini:1969qy}. Later Colpi and collaborators
 generalized these solutions by including a 
self-interacting scalar field \cite{Colpi}. A review of their main properties can be found in the following references \cite{Jetzer:1991jr,Schunck:2003kk,Liebling:2012fv}. Attempts to generalize the spherically symmetric solutions to include rotation were performed by Kobayashi, Kasai and Futusame who investigated the slow rotation limit of BSs perturbatively \cite{Kobayashi:1994qi}. 
However, they failed in doing so because they 
did not consider a sufficiently general harmonic ansatz compatible with axisymmetry, and concluded that the slow rotation approximation was not possible. Other authors 
even questioned the possibility of finding full non-linear {\it fast rotating BSs}\cite{Yoshida1997a}. Notwithstanding, 
the fully rotating BS analysis was described first by Schunck and Mielke in the weakly relativistic regime \cite{Schunck:1996he} and by Yoshida and Eriguchi in the highly relativistic regime \cite{Yoshida:1997qf}. Further studies, including higher winding number solutions 
(i.e. excited BS solutions that correspond to scalar-field solutions  with {\it nodes}) were found in \cite{Grandclement:2014msa, Ontanon:2021hbg}. 

From the astrophysical point of view, BSs can play an important role in several scenarios. For instance, they have been considered as an alternative to standard structure formation through Dark Matter seeds \cite{Matos:1998vk,Matos:1999et,Matos:2000ss,Hui:2016ltb,Mielke:2000mh,Klaer:2017ond} or as black hole mimickers \cite{Schunck:1997dn,Torres:2000dw,Lemos:2008cv,Mazur:2001fv,Guzman:2005bs,Vincent:2015xta,Meliani:2015zta}. Recently, it has been shown that the gravitational wave signal of the collision of two extreme compact objects can be described in terms of the collision of two vector BSs \cite{Sanchis-Gual:2018oui}.

The study of rotating BSs is relevant because they present some distinctive properties, namely, the
energy-density is off the geometrical center and it is distributed in a torus shape, the ratio between its angular momentum and the number of particles is an integer, and in general, the dynamics of particles (geodesics) around rotating BSs is richer than the static counterpart in the sense that new regions of stability appear \cite{Grandclement:2016eng,Zhang:2021xhp}.
Furthermore, the photons moving around a BSs may lead to a shadow consistent with the one reported by the Event Horizon Telescope \cite{Olivares:2018abq}.

In this paper we revisit the rotating BSs scenario by considering the slow rotation approximation. 
The advantage of this approach is that one is left with a system of ordinary differential equations for the metric that are simpler to solve than the fully rotating case which consists of partial differential equations (PDEs), and the approximate solutions capture some of the main properties of the fully rotating case.

In the present work, we present sequences of rotating BS solutions and describe some of their properties. We make a thorough comparison of some physical quantities between the solutions in the slow rotation approximation and the solutions in the fully rotating case, and thus estimate some regions of the parameters where the approximation is valid.

The paper is organized as follows: In Section \ref{sec:framework} we present the basic setup, the field equations and some considerations to find equilibrium configurations. In Section \ref{sec:SRBS} we describe the slow rotation approximation and  explicitly display the equations of motion. We also discuss the viability of the approach. In Section \ref{sec:results} we present the numerical results and a detailed description of the solutions along some physical quantities. We further describe the geodesic motion of particles moving in the vicinity of BSs by analyzing the effective potentials and the structure of circular orbits. Finally in Section \ref{sec:final} we give some concluding remarks.

Throughout this work we use units where $G = c = \hbar = 1$ and use the metric signature ($-$, $+$, $+$, $+$).

\section{Framework and formalism}
\label{sec:framework}

We consider the Einstein-Klein-Gordon (EKG) field theory with a complex scalar field $\Psi$ with massive but 
free field minimally coupled to Einstein gravity. The model is described by the action
\begin{equation}\label{Eq:Action}
	\mathcal{S} = \int d^4 x \sqrt{-g} \left[ \frac{\mathcal{R}}{16\pi} - g^{\alpha\beta} \partial_\alpha \Psi^* \partial_\beta \Psi - \mu^2 \Psi^* \Psi \right] ~,
\end{equation}
where $\mathcal{R}$ is the Ricci scalar, $\mu$ is the mass of the scalar field and the asterisk $(*)$ denotes the complex conjugate.
By performing the variation of the above action with respect to the metric and the scalar field, one obtains the equations of motion,
\begin{eqnarray}
	E_{\alpha\beta} \equiv  R_{\alpha\beta} - \frac{1}{2} g_{\alpha\beta} \mathcal{R} - 8\pi T_{\alpha\beta} &=& 0 ~,	\label{Eq:EinsteinFieldEquations}\\
	\Box \Psi - \mu^2 \Psi &=& 0 ~, \label{Eq:KleinGordonEquation}
\end{eqnarray}
where $\Box$ is the covariant d'Alembert operator, $\Box \equiv \frac{1}{\sqrt{-g}} \partial_\alpha \left( \sqrt{-g} g^{\alpha\beta}\partial_\beta \right)$, and $T_{\alpha\beta}$ is the energy-momentum tensor (EMT) associated with the scalar field defined as,
\begin{equation}
	T_{\alpha\beta} = 2 \partial_{(\alpha} \Psi^*  \partial_{\beta)} \Psi - g_{\alpha\beta} \left( \partial^\gamma \Psi^* \partial_\gamma \Psi + \mu^2 \Psi^* \Psi \right) ~.
\end{equation}

A generic feature of this theory is the existence of a $U(1)$ global symmetry generated by the scalar field transformation $\Psi \rightarrow e^{i \sigma} \Psi$, where $\sigma$ is a constant that leaves the action, Eq.~\eqref{Eq:Action}, invariant. Noether's theorem 
associated with this symmetry leads to a conserved current that, in turn, provides a corresponding conserved Noether charge, which is interpreted as the total number of bosons. The conserved current can be written as $j^\alpha = -i \left( \Psi^* \partial^\alpha \Psi - \Psi \partial^\alpha \Psi^* \right)$ whose local  conservation $\nabla_\alpha j^\alpha = 0$ can be verified.
The Noether charge can be computed through the integration of the time-like component of the conserved current,
\begin{equation}
	Q = \int_\Sigma j^t ~,
\end{equation} 
where $\Sigma$ is a spacelike hypersurface bounded by a 2-sphere at spatial infinity. As we stressed above, we can interpret this quantity as the total number of scalar particles of a solution.
Moreover, 
in the axisymmetric rotating scenario within the theory in consideration, it is possible to show the existence of a direct relation between the Noether charge, $Q$, and the total angular momentum, $J$, of the BS
\begin{equation}
	J = m Q ~,
\end{equation}
where $m$ is an integer, and, as we show below, it corresponds to the azimuthal harmonic index of the scalar field that is associated with the axial symmetry of the rotating BS.

\subsection{Stationary equilibrium configurations: 
Axisymmetric Rotating Boson Stars}

We are interested in solutions of the Einstein-Klein-Gordon system of Eqs. \eqref{Eq:EinsteinFieldEquations}
and \eqref{Eq:KleinGordonEquation}, that are horizonless, regular everywhere, stationary, axisymmetric and asymptotically flat. Therefore, a possible metric \textit{ansatz} that reflect such conditions 
and compatible with the absence of meridional currents (i.e. the circularity condition \cite{Wald})
is,
\begin{equation}\label{Eq:AnsatzMetricFull}
	ds^2 = -e^{2F_0(r,\theta)} dt^2 +  e^{2F_1(r,\theta)} \left( dr^2 + r^2 d\theta^2 \right) + e^{2F_2(r,\theta)}r^2 \sin^2 \theta \left(d\varphi - \frac{W(r,\theta)}{r} dt \right)^2  ~,
\end{equation}
For the scalar field, we consider the stationary and axisymmetric form,
\begin{equation}\label{Eq:Ansatzsfield}
	\Psi(t,r,\theta,\varphi) = \phi(r,\theta) e^{i(\omega t - m \varphi)} ~, 
\end{equation}
where $\omega$ and $m = \pm 1, \pm 2, \dots$ are the frequency and azimuthal harmonic index of the scalar field, respectively. 
This ansatz leads to an EMT that 
is compatible with the underlying symmetries of the spacetime: stationarity and axial symmetry. 
Moreover, this ansatz allows us to consider the 
slow rotation approximation that is only possible 
when $m\neq 0$. As shown in Ref.\cite{Kobayashi:1994qi} when $m=0$, slowly rotating BS solutions around the spherically symmetric BS configurations are not possible.

The metric and scalar field presented above are well motivated and commonly adapted in the literature regarding fully rotating BSs. Furthermore, these features have been implemented, for instance, in studies of BSs with different scalar-field potentials~\cite{Herdeiro:2016gxs,Delgado:2020udb}, 
hairy rotating black holes
with higher azimuthal harmonic index~\cite{Delgado:2019prc}, gauged BSs~\cite{Delgado:2016jxq,Herdeiro:2021jgc}, multistate BSs~\cite{Li:2020ffy,Zeng:2023hvq} and universal relations associated with BSs~\cite{Adam:2022nlq}.

\section{Slowly rotating boson stars}
\label{sec:SRBS}

In order to construct solutions in the slow rotation regime we assume an spacetime metric of the form:
\begin{equation}\label{Eq:AnsatzMetric}
	ds^2 = -e^{2F_0(r)} dt^2 +  e^{2F_1(r)} \left[ dr^2 + r^2 d\theta^2 + r^2 \sin^2 \theta \left(d\varphi - \frac{W(r)}{r} dt \right)^2 \right] ~,
\end{equation}
where, in contrast with the spacetime \eqref{Eq:AnsatzMetricFull}, the functions $F_0$, $F_1$ and $W$ depend only on $r$ and $F_2=F_1$. The metric \eqref{Eq:AnsatzMetric} represents small deviations from a spherically symmetric spacetime. 
This form for the spacetime metric 
was considered in the past (mutatis mutandis) when analyzing slowly rotating neutron stars modeled with a perfect fluid\cite{Hartle}.
Given the metric \eqref{Eq:AnsatzMetric} and the scalar field \eqref{Eq:Ansatzsfield}, one can write explicitly the equations of motion. Since we will focus on slowly rotating BSs, we consider only the first order (linear) corrections in $W(r)$ and neglect higher order terms. The Einstein field equations yield,
{\small
\begin{eqnarray*}
	E_t^t = 0 &\Rightarrow& 2 F_1'' + \frac{4}{r} F_1' + F_1'^2 + 8\pi \left[ e^{2F_1} \phi^2 \left( \mu^2 + e^{-2F_0} \omega^2 + e^{-2F_1} \frac{m^2}{r^2 \sin^2 \theta} \right) + \frac{(\partial_\theta \phi)^2}{r^2} + (\partial_r \phi)^2 \right] = 0~, \\
	E_r^r = 0 &\Rightarrow& F_1'^2 + 2 F_0' F_1' + \frac{2}{r} \left(F_1' + F_0' \right) + 8\pi \left\{ e^{2F_1} \phi^2 \left[ \mu^2 - e^{-2F_0} \omega \left( \omega - 2 m \frac{W}{r} \right) + e^{-2F_1} \frac{m^2}{r^2 \sin^2 \theta} \right]  + \frac{(\partial_\theta \phi)^2}{r^2} - (\partial_r \phi)^2\right\} = 0~,\\
	E_\varphi^t = 0 &\Rightarrow& W'' + \left(W' - \frac{W}{r} \right) \left(\frac{2}{r} - F_0' + 3F_1' \right) + \frac{32\pi m}{r \sin^2 \theta} \phi^2 \left(\omega - m \frac{W}{r} \right) = 0~, \\
 E_t^\varphi = 0 &\Rightarrow&  E_\varphi^t + 2 W \left[ 2 8\pi \phi^2 \left( e^{-2F_0 + 2F_1} \omega^2 + \frac{m^2}{r^2 \sin^2\theta}\right) - \frac{F_0'}{r} - F_0'^2 - F_0'' + \frac{3F_1'}{r} + F_1'^2 + F_1'' \right] = 0 \\
	E_r^\theta = 0 &\Rightarrow& - 16\pi \frac{e^{-2F_1}}{r^2} \partial_r \phi \partial_\theta \phi = 0~, \\
	E_\theta^r = 0 &\Rightarrow& r^2 E_r^\theta = 0~, \\
	E_\theta^\theta = 0 &\Rightarrow& F_1'' + F_0'' + F_0'^2  + \frac{F_0'}{r} + \frac{F_1'}{r} + 8\pi \left\{ e^{2F_1} \phi^2 \left[ \mu^2 - e^{-2F_0} \omega \left( \omega - 2 m \frac{W}{r} \right) + e^{-2F_1} \frac{m^2}{r^2 \sin^2 \theta} \right] - \frac{(\partial_\theta \phi)^2}{r^2} + (\partial_r \phi)^2\right\} = 0~, \\
	E_\varphi^\varphi = 0 &\Rightarrow& E_\theta^\theta - 16\pi \left[ e^{2F_1} \phi^2 \left( e^{-2F_0} m \omega \frac{W}{r} + e^{-2F_1} \frac{m^2}{r^2 \sin^2 \theta} \right) - \frac{(\partial_\theta \phi)^2}{r^2} \right] = 0~,
\end{eqnarray*}
}
where the prime denotes the radial derivative. 
At this point we are still keeping the angular-$\theta$ and radial dependency of the 
scalar field in the EMT. Later, and for consistency, we plan to average the Einstein equations on the $\theta$ coordinate, while keeping this angular dependency in the Klein-Gordon equation.
The Klein-Gordon equation yields,
\begin{equation}
	r^2 \left\{ \partial_r^2 \phi + \partial_r \phi \left(\frac{2}{r} + F_0' + F_1' \right) + e^{-2F_1} \phi \left[ -\mu^2 + e^{2F_0} \omega \left( \omega - 2 m \frac{W}{r} \right) \right] \right\} = \partial_\theta^2 \phi + \cot \theta \partial_\theta \phi - \frac{m^2 }{\sin^2 \theta} \phi~. 
\end{equation}
By writing the scalar field as $\phi = R(r) \Theta(\theta)$, the previous equation can be separated as follows
\begin{eqnarray}
	R'' + R' \left(\frac{2}{r} + F_0' + F_1' \right) + e^{2F_1} R \left[ -\mu^2 + e^{-2F_0} \omega \left( \omega - 2 m \frac{W}{r} \right) \right] &=&  \ell (\ell + 1) \frac{R}{r^2} \label{Eq:RadialKleinGordonEquation} \\
	\partial_\theta^2{\Theta} + \cot \theta \partial_\theta{\Theta} - \frac{m^2}{\sin^2 \theta} \Theta &=&  -\ell(\ell + 1)  \Theta ~, \label{Eq:SphericalHarmonicsEquation}
\end{eqnarray}
where
$\ell$ is a separation constant that turns to be a non-negative integer in order to ensure the regularity on the 
axis of symmetry $\theta=\{0,\pi\}$, in a similar way as in the 
analysis of the Schr\"odinger equation for the Hydrogen atom. In fact, the solution of \eqref{Eq:SphericalHarmonicsEquation} together with the azimuthal dependency $\varphi$ is given in terms of the spherical harmonics that are characterized by the integer numbers $\ell$ and $m$ with 
$|m|\leq \ell.$

\subsection{Asymptotically flat and regular solutions}
If we take into account the angular-$\theta$ dependency of the scalar field in the EMT we conclude
from the components $E_r^\theta = E_\theta^r = 0$ and the separation of the radial and angular parts of the scalar field, a nontrivial radial distribution of scalar field must satisfy
$\partial_\theta \Theta = 0$, which, by means of Eq. \eqref{Eq:SphericalHarmonicsEquation}, implies that the only possible solution is the spherical harmonic with 
$\ell = m = 0$,
\begin{equation}
	\Theta = \frac{1}{2\sqrt{\pi}} ~.
\end{equation}
Using this result, one can easily see that the $E_\varphi^t = 0$ equation yields,
\begin{equation}
	W'' + \left( W' - \frac{W}{r} \right) \left( \frac{2}{r} - F_0' + 3 F_1' \right) = 0 ~.
\end{equation}
If we introduce a new function, $\overline{W} = r W$, the above equation simplifies to,
\begin{equation}
	\overline{W}'' + \overline{W}' \left(\frac{4}{r} - F_0' + 3 F_1' \right) = 0 ~.
\end{equation}
This equation is equivalent to Eq. (4.4) of~\cite{Kobayashi:1994qi} with $q = C_0 = a_1 = 0$.
After a direct integration one gets,
\begin{equation}
	e^{3F_1 + F_0} r^4 \overline{W}' = \text{const} ~.
\end{equation}
Due to the requirement of asymptotic flatness, the constant must vanish and $\overline{W} = W =0$.
Hence, the system of equation reduces to the spherically symmetric scenario and no slowly rotating BSs seem possible. This conclusion is consistent with the results presented earlier by 
Kobayashi, Kasai and Futusame in~\cite{Kobayashi:1994qi}. 
Nevertheless, as we analyze in the next section, we can overcome this drawback by considering an approximation for the backreaction of the angular contribution $\theta$ of the scalar field in the spacetime only in the average sense. This approximation will keep the Einstein equations mathematically consistent while keeping all the angular dependency $\theta$ and $\varphi$
in the scalar field intact. At this respect we remind the 
reader that it is common to take similar approximations where, actually, non-trivial scalar field solutions 
are considered when the field is taken as a 
{\it test field} that does not backreact into a given spacetime background. Here instead, we consider that most of the field backreacts fully into the spacetime except for 
the $\theta$ dependency that is averaged out.

\subsection{Approximate averaged solutions}

As shown in the previous 
section, under the given assumptions, it seems that no regular and asymptotically flat solutions of the Einstein-Klein-Gordon system are possible in the slow rotation regime.
However, it is possible to find rotating solutions by considering the average on the angular $\theta$ part 
in the Einstein equations as we describe below.
Let us start first by discussing some properties of the system of equations. 

The system \eqref{Eq:EinsteinFieldEquations}
is invariant under a scaling with  
$\mu$ by making the substitutions, 
\begin{equation}
	r \rightarrow r \mu ~, \hspace{10pt} \omega \rightarrow \omega/\mu ~.
\end{equation}
As a consequence, the global quantities, such as the ADM mass and angular momentum, will be expressed in units of $\mu$.
From the metric,~\eqref{Eq:AnsatzMetric}, and the decomposition of the scalar field, $\phi(r,\theta) = R(r)\Theta(\theta)$, one is left with 4 unknown functions, $F_0$, $F_1$, $W$ and $R$ and the unknown frequency $\omega$. These quantities will be found 
after imposing an appropriate set of boundary conditions.
For convenience we also rescale the scalar field as $\phi \rightarrow \phi \sqrt{4\pi}$. 
Furthermore, in order for the system of equations to be solved explicitly, one has to provide specific values 
for $\ell$ and $m$. 
By solving the system of equations one has a family of solutions that is also characterized by the number of radial nodes, $n$, in $R(r)$. In the present work, we will focus on fundamental configurations, i.e. configurations with a nodeless scalar field, $n=0$.

A common approach to solve the Einstein's equations in the stationary and axisymmetric scenario is to consider a suitable combination of them.
This method has been implemented in the past to find BS and hairy black holes solutions \cite{Herdeiro:2014goa,Herdeiro:2015gia,Delgado:2016jxq,Delgado:2020udb,Delgado:2020hwr}. The idea relies on finding a combination of the Einstein equations that gives second order radial derivatives of a single unknown function. The linear combination of the equations, together with the Klein-Gordon, Eq. \eqref{Eq:RadialKleinGordonEquation}, that we consider are as follows,
{\small
\begin{eqnarray}
	E_t^t + E_r^r - E_\theta^\theta + E_\varphi^\varphi = 0 &\Rightarrow& F_1'' + F_1'^2 + F_0' F_1' + \frac{3}{r} F_1' + \frac{F_0'}{r} + 2 R^2 \left( e^{2F_1} \Theta^2 + \frac{2}{r^2} \left(\partial_\theta \Theta \right)^2  \right) = 0 ~, \\
	-E_t^t + E_r^r + E_\theta^\theta + E_\varphi^\varphi = 0 &\Rightarrow& F_0'' + F_0'^2 + F_0' F_1' + \frac{2}{r} F_0' + 2 e^{2F_1} R^2 \Theta^2 \left[ 1 + 2 e^{-2F_0} \omega \left( \omega - m \frac{W}{r} \right) \right] = 0 ~, \\
	E_\varphi^t = 0 &\Rightarrow& W'' + \left( W' - \frac{W}{r} \right) \left( \frac{2}{r} - F_0' + 3 F_1' \right) + \frac{8 m}{r \sin^2 \theta} R^2 \Theta^2 \left( \omega - m \frac{W}{r} \right) = 0 ~, \\
	\Box \Psi - \mu^2 \Psi = 0 &\Rightarrow& R'' + R' \left( \frac{2}{r} + F_0' + F_1' \right) - e^{2F_1} R \left[ 1 - e^{-2F_0} \omega \left( \omega - 2 m \frac{W}{r} \right) + \frac{\ell (\ell + 1)}{r^2} \right] = 0 ~.
\end{eqnarray} }
Since we only used 3 of the 6 independent Einstein equations, we can use the remaining 3 equations, $E_t^t - E_r^r - E_\theta^\theta + E_\varphi^\varphi = 0$, $-E_r^r + E_\theta^\theta = 0$ and $E_r^\theta = 0$, as ``constraints", 
as a way to monitor the accuracy of the numerical solutions.

We solve the system of equations under the following boundary conditions for the unknown functions. Asymptotically we impose that the spacetime is flat which imply,
\begin{equation}
	\text{At }~r \rightarrow \infty ~: \hspace{10pt} F_1 = F_0 = W = R = 0 ~.
\end{equation}
On the other end, i.e., at the origin and on the axis of symmetry, we impose that the solutions are regular there. By performing a power expansion of the 
functions around $r=0$, we find that the differential equations imply the following conditions at the origin,\footnote{The behavior  $R\sim r^{\ell}$ is consistent with the boundary conditions and has been used previously in \cite{Alcubierre:2021psa} to find regular solutions in spherical symmetry.}
\begin{equation}
	\text{At }~r \rightarrow 0 ~: \hspace{10pt} \partial_r F_1 = \partial_r F_0 = W = R = 0 ~.
\end{equation}
Notice that in the rotating case, 
the regularity conditions on the origin with $\ell\neq 0$
implies that the topology of the scalar field is toroidal. Notably, the 
scalar field vanishes at the origin while its radial derivative $R'(0)\neq 0$, precisely the opposite what happens in the the spherically symmetric scenario where $R(0)\neq 0$ and $R'(0)=0$.

The strategy to obtain regular BS solutions in the slow rotation approximation is to average the Einstein equations over the solid angle, $d\Omega = \sin \theta d\theta d\varphi$. In other words, for each of the above 
Einstein equations, $\mathcal{E}$, we perform the following average,
\begin{equation}\label{Eq:WeightedAverage}
	\tilde{\mathcal{E}} = \frac{\displaystyle\int \mathcal{E} d\Omega }{\displaystyle\int d\Omega} = \frac{\displaystyle \int_0^{2\pi}  d\varphi \int_0^\pi d\theta \sin \theta\ \mathcal{E} }{\displaystyle \int_0^{2\pi} d\varphi \int_0^\pi d\theta \sin \theta} =  \frac{1}{2} \int_0^\pi d\theta \sin \theta\ \mathcal{E}  ~,
\end{equation}
where $\tilde{\mathcal{E}}$ denote the averaged equation of motion. The advantage of this approach is two-folded. First, any dependency on the angle $\theta$ disappears, and thus, we only need to solve a system of ordinary differential equations (ODEs), instead of a system of PDEs. Second, and more importantly, 
considering that the solutions of 
\eqref{Eq:SphericalHarmonicsEquation} (together with the $\varphi$ dependence)
are the spherical harmonics, 
equation $E_r^\theta = 0$ will always be satisfied on average. 
In the present context, we will focus only on the case where $l=m=1$ 
(modes with $m = - 1$ have basically the same properties):
\begin{equation}\label{Eq:AngularPartScalarField}
	\Theta(\theta) = - \frac{1}{2} \sqrt{\frac{3}{2\pi}} \sin \theta ~.
\end{equation}
The final system of ODEs that we are interested to solve simplifies to,
{\small
\begin{eqnarray}
 	 \tilde{E_t^t} + \tilde{E_r^r} - \tilde{E_\theta^\theta} + \tilde{E_\varphi^\varphi} = 0 &\Rightarrow& F_1'' + F_1'^2 + F_0' F_1' + \frac{3}{r} F_1' + \frac{F_0'}{r} + \frac{1}{2\pi} R^2 \left( e^{2F_1} + \frac{1}{r^2}  \right) = 0 ~, \\
	-\tilde{E_t^t} + \tilde{E_r^r} + \tilde{E_\theta^\theta} + \tilde{E_\varphi^\varphi} = 0 &\Rightarrow& F_0'' + F_0'^2 + F_0' F_1' + \frac{2}{r} F_0' + \frac{1}{2\pi} e^{2F_1} R^2 \left[ 1 + 2 e^{-2F_0} \omega \left( \omega - \frac{W}{r} \right) \right] = 0 ~, \\
	\tilde{E_\varphi^t} = 0 &\Rightarrow& W'' + \left( W' - \frac{W}{r} \right) \left( \frac{2}{r} - F_0' + 3 F_1' \right) + \frac{3}{\pi r} R^2 \left( \omega - \frac{W}{r} \right) = 0 ~, \\
	\tilde{\Box \Psi} - \tilde{\mu^2 \Psi} = 0 &\Rightarrow& R'' + R' \left( \frac{2}{r} + F_0' + F_1' \right) - e^{2F_1} R \left[ 1 - e^{-2F_0} \omega \left( \omega - 2 \frac{W}{r} \right) + \frac{2}{r^2} \right] = 0 ~.
\end{eqnarray} }
The numerical scheme to solve the above system of ODEs is as follows. First, we define a new radial coordinate, $x$, that allow us to map the semi-infinite region $[0,\infty)$ to the finite region $[0,1]$. This is done by defining such coordinate as, $x =  r/(r+1)$. Second, we discretize the above system of ODEs on a radial grid with 801 points. Finally, we solve the system with the aforementioned boundary conditions using the professional package \texttt{FIDISOL/CADSOL}. This package is based on a finite-difference approach together with a Newton-Raphson method where the user can use an arbitrary grid and arbitrary consistency order. 
It also provides an error estimate for each numerical solution. 
For all the solutions presented here, the maximal numerical relative error is of the order of $10^{-5}$.
However, since we know these are solutions of a system of equations that was constructed by an average approximation, we expect that the average version of the constraint equations, $\tilde{E}_t^t - \tilde{E}_r^r - \tilde{E}_\theta^\theta + \tilde{E}_\varphi^\varphi = 0$ and $-\tilde{E}_r^r + \tilde{E}_\theta^\theta = 0$ will not be fully satisfied 
pointwise but only on average.
Indeed, we verified that only on the Newtonian regime, where the solutions are very diluted and the frequency $\omega$ is close to unity, and the scalar field is small, the averaged constraint equations yield an error smaller than $10^{-5}$. Nevertheless, as we will show, the approximate average solutions can yield values for the physical quantities that are in agreement with the ones for the fully rotating BSs, even for solutions outside of the Newtonian regime. 
On this note, we will consider the
level of agreement with the fully rotating case as the range of validity of the slow rotation approximation.

\section{Numerical results}
\label{sec:results}

\begin{figure}[ht!]
	\centering
	\includegraphics[width=0.49\linewidth]{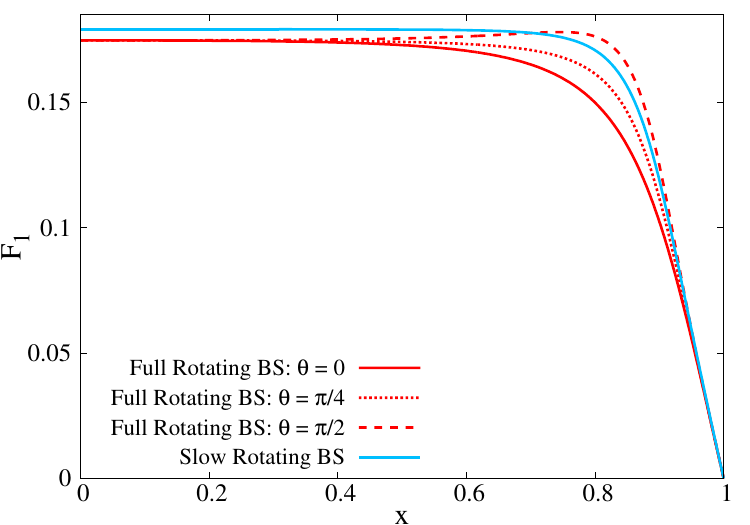}
	\includegraphics[width=0.49\linewidth]{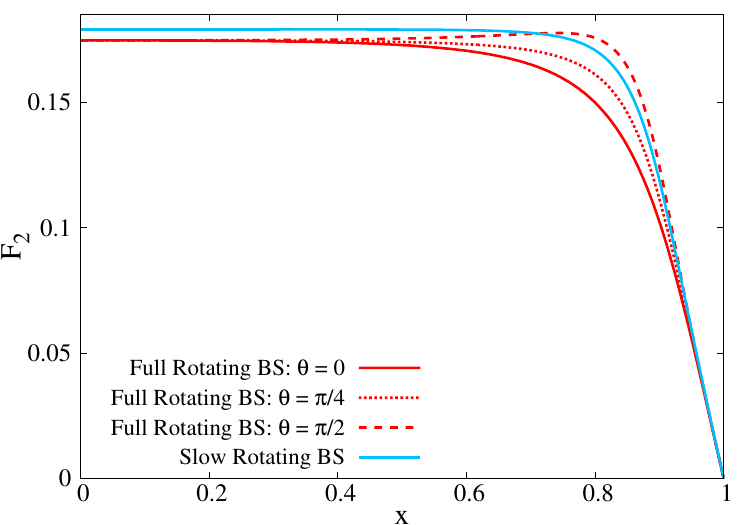}\\
    \includegraphics[width=0.49\linewidth]{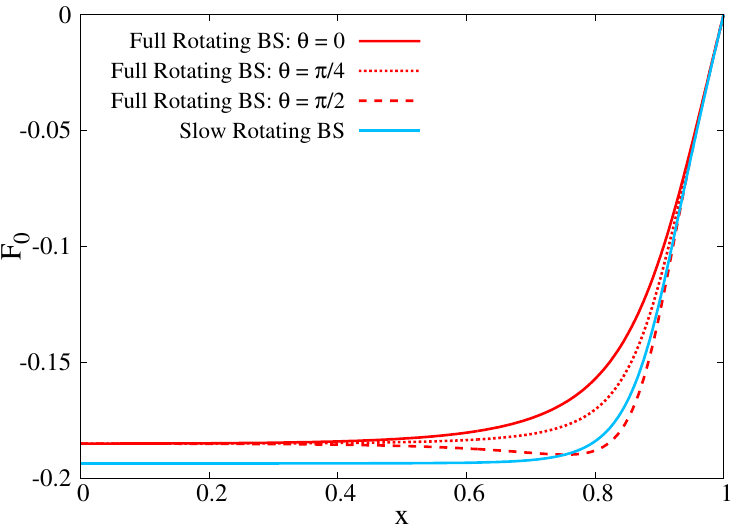}
	\includegraphics[width=0.49\linewidth]{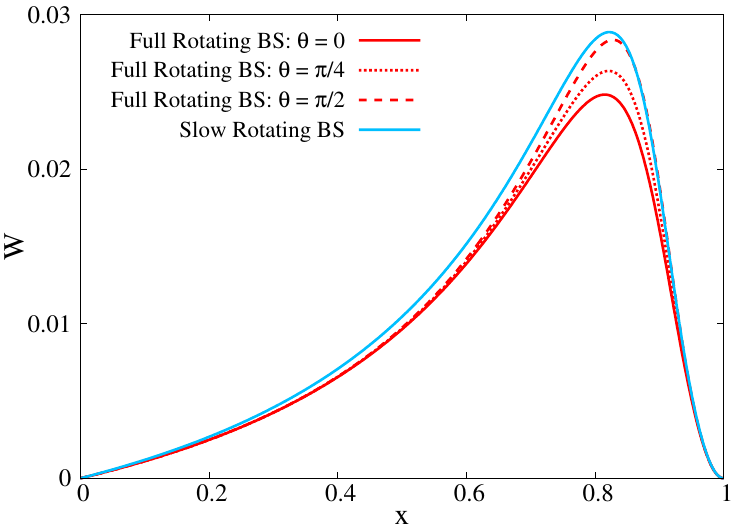}\\
    \includegraphics[width=0.49\linewidth]{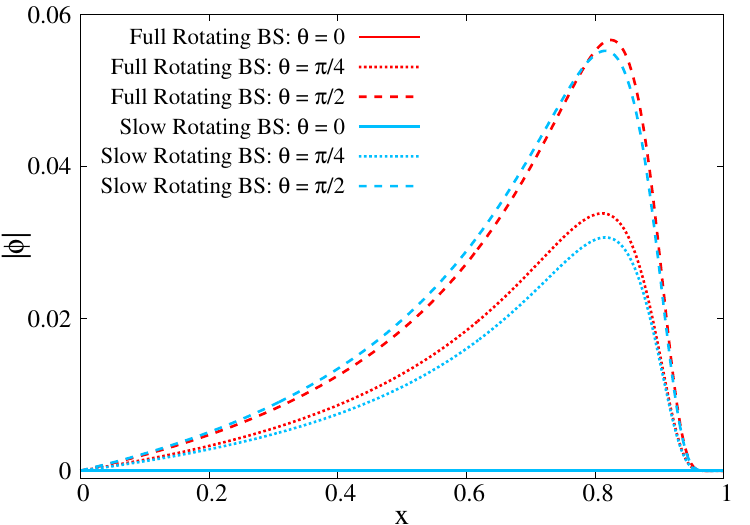}
	\caption{Profile of the metric and scalar field functions of a typical fully rotating BS (red lines) and a slowly rotating BS (blue lines) with $\omega/\mu = 0.9$, as a function of the compact radial coordinate, $x = r/(r+1)$. The three line styles correspond to the \textit{ansatz} function evaluated at a specific angular coordinate. Note that, for slowly rotating BS, $F_2 = F_1$, and  $F_1, F_0$ and $W$ do not depend on the angular coordinate, therefore only one line is presented for each of them.}
	\label{Fig:AnsatzFunctions}
\end{figure}

Using the aforementioned solver, we obtain close to a thousand numerical solutions by varying the only input parameter available, the angular frequency of the scalar field, $\omega$. For all the solutions obtained, the functions $F_1, F_0, W$ and $R$ have smooth profiles, as well as their first and second radial derivatives, which yield finite curvature invariants on the full domain of integration.  
To motivate this claim, the profiles of the functions of a typical full and slowly rotating BS as a function of the compact radial coordinate, $x = r/(r+1)$, are presented in Fig. \ref{Fig:AnsatzFunctions}. We have chosen a BS with $\omega/\mu = 0.9$. In Fig. \ref{Fig:AnsatzFunctions}, three red lines are shown corresponding to each function of the fully rotating BS evaluated at three different angular values, $\theta = \{0, \pi/4, \pi/2\}$. For the $F_1$, $F_0$ and $W$ profiles, only one blue line is shown, which corresponds to its profile for the slowly rotating BS, since these profiles do not depend on the angular coordinate $\theta$. For the scalar field profile, however, it depends on the angular coordinate, hence, three blue lines are plotted for the same previous three angular values. Furthermore, note that, as mentioned in the beginning of the previous section, $F_1 = F_2$ for the slowly rotating BSs. 
We stress again that 
here we are interested only in the modes with $\ell=|m|=1$, for both the slow rotation approximation, and the fully rotating BS. 

\subsection{Physical quantities}

\begin{figure}[ht!]
	\centering
	\includegraphics[width=0.65\linewidth]{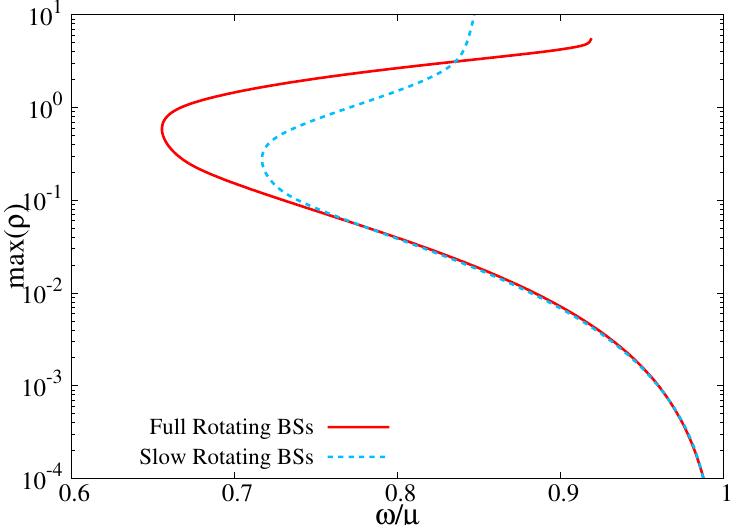}
    \caption{Maximum of the energy-density measured by a ZAMO observer, $\text{max}(\rho)$, as a function of the angular frequency of the scalar field, $\omega/\mu$, for fully rotating (red solid line) and slowly rotating BSs (blue dashed line).}
	\label{Fig:maxrho_omega}
\end{figure}

The first physical quantity that we focus on is the energy-density, $\rho$. This quantity is highly important since it is directly related with the energy conditions, in particular, with the weak energy condition, that states that for any time-like vector, $X^\mu$, the energy-density measured by the corresponding observer must be non-negative. Mathematically speaking, $\rho = T_{\mu\nu} X^\mu X^\nu \geq 0$. 
For simplicity, in this work, we choose a zero angular momentum observer (ZAMO), defined as an observer whose 4-velocity is orthogonal to $t = \text{constant}$ hyper-surfaces, 
{
$X^a= X^t t^a + X^\varphi \varphi^a $, for suitable $X^t$ and $X^\varphi$, where $t^a=\left(\partial_t \right)^a$ and $\varphi^a=\left(\partial_\varphi\right)^a$ are the timelike and rotational Killing vector fields, respectively, which are associated with the stationary and axial symmetries of the spacetime considered here\footnote{In the framework of the 3+1 Formalism of GR, these componentes are given in terms of the {\it lapse} function $N$ and the shift vector $N^\varphi$: $X^t=1/N= e^{-F_0}$ and $X^\varphi=N^\varphi/N= We^{-F_0}/r$.}.
}
For such observer~\cite{Delgado:2020udb},
\begin{equation}
	\rho = - T_t^t - T_\varphi^t \frac{g^{\varphi t}}{g^{tt}} \geq 0 ~.
\end{equation}
In Fig. \ref{Fig:maxrho_omega}, we present how the maximum value of the energy-density, $\text{max}(\rho)$ varies by changing the angular frequency, $\omega/\mu$, of both fully rotating (red solid line) and slowly rotating BSs (blue dashed line). 
In the Newtonian limit, where $\omega/\mu \sim 1$, all solutions have very small energy-densities, $\text{max}(\rho) < 10^{-4}$. As we go along both lines, the angular frequency first decreases and, after the backbend where the first branch ends and the second initiates, it starts to increase again. The energy-density {$\text{max}(\rho)$}, however, is a monotonically increasing quantity. As we approach the strong gravity regime, the energy-density of the solutions increases.

In this plot, we can verify that both lines are in good agreement for all solutions with $\omega \gtrsim 0.75$ and $\text{max}(\rho) \lesssim 10^{-1}$. This is expected since, in such region, $W \ll 1$, and the linear approximation holds. However, outside of this region, it is possible to see large differences.
The biggest one resides on the minimal value of the frequency for both lines. The fully rotating BSs can have a minimal frequency of $\omega/\mu \sim 0.65$, whereas the slowly rotating BSs can only go down to $\omega/\mu \sim 0.72$. The same happens as we go along the lines (increasing the maximum of the energy-density) until the last solution. Here, for fully rotating BSs, $\omega/\mu \sim 0.92$, and after that, the angular frequency start increasing again, creating a third branch of solutions~\cite{Herdeiro:2015gia}. In contrast, for slowly rotating BSs, the last solution has $\omega/\mu \sim 0.85$ and we could not find a third branch, as in the fully rotating counterpart. 
These differences arise because $W$ is no longer small enough for the linear approximation to hold, and, as $W$ increases, the metric functions have a significant angular dependency which we are neglecting in this approximation.
Two of the most important physical quantities that characterize BSs are the ADM mass $M$, and the angular momentum $J$. These can be obtained through the asymptotic behaviour of the following two metric functions,
\begin{equation}
	g_{tt} = -e^{2F_0} + e^{2F_1} W^2 \sin^2 \theta = - \left( 1 - \frac{2M}{r} \right) + \dots ~, \hspace{20pt} g_{t\varphi} = - e^{2F_1} W r \sin^2 \theta = -\frac{2J}{r} \sin^2 \theta + \dots ~.
\end{equation}
Alternatively, one can also compute such quantities through the corresponding Komar integrals \cite{Poisson_2004},
\begin{equation}
	M = - 2 \int_\Sigma d S_\mu \left( T_\nu^\mu  t^\nu - \frac{1}{2} T t^\mu \right) ~, \hspace{20pt} J = \int_\Sigma d S_\mu \left( T_\nu^\mu \varphi^\nu - \frac{1}{2} T \varphi^\mu \right) ~,
\end{equation}
where $\Sigma$ is a spacelike surface bounded by a 2-sphere at spatial infinity, $S_\infty$, and as we mentioned before, $t^\mu$ and $\varphi^\mu$ are the timelike and rotational Killing vector fields, respectively. These integrals amount to the mass and angular momentum stored in the 
boson star,
and, since the solutions are entirely composed of a scalar field and are asymptotically flat, then the mass and angular momentum computed by the two aforementioned ways must yield the same results. Therefore, a good numerical test consists in comparing the mass and angular momentum obtained by the asymptotic behaviour of the metric with the Komar integrals and verify their equality. For all slowly rotating BSs the maximal deviation observed was of the order of $10^{-4}$. 

\begin{figure}[ht!]
	\centering
	\includegraphics[width=0.49\linewidth]{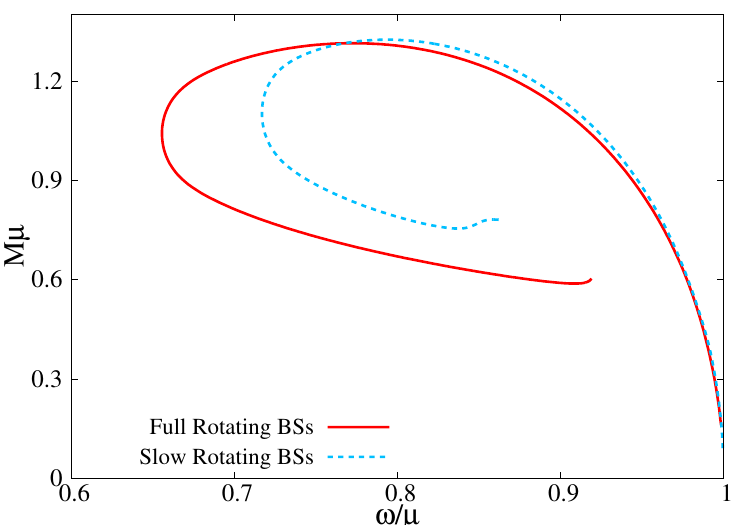}
	\includegraphics[width=0.49\linewidth]{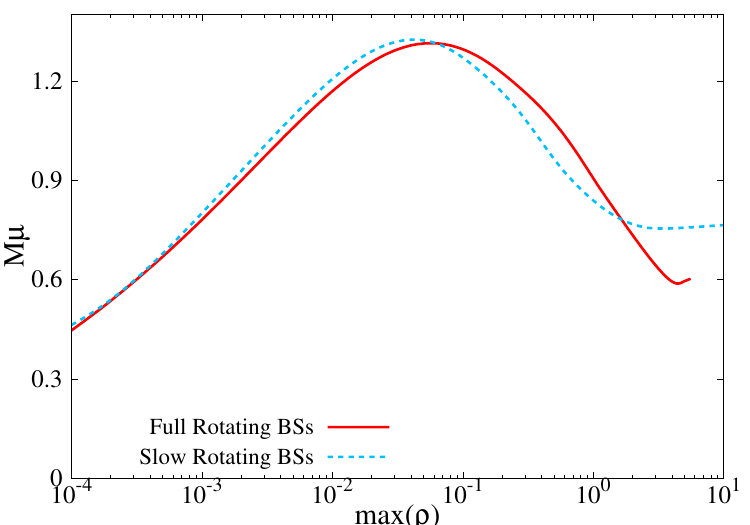}
	\caption{ADM mass, $M\mu$, as a function of the angular frequency, $\omega/\mu$, (left panel) and maximum of the energy-density, $\text{max}(\rho)$, (right panel), for fully rotating BSs (red solid line) and slowly rotating BSs (blue dashed line).}
	\label{Fig:Mass_omega_maxrho}
\end{figure}

In Fig. \ref{Fig:Mass_omega_maxrho}, it is shown the ADM mass of the fully rotating BSs (solid red line) and slowly rotating BSs (blue dashed line) as a function of both the angular frequency (left panel) and the maximum of the energy-density (right panel). 
In the left panel, one can compare solutions with the same angular frequency. Here we see that the ADM mass of fully rotating BSs can be well approximated by the slowly rotating BSs for $\omega/\mu \gtrsim 0.75$ in the first branch. The slowly rotating BSs present slightly larger ADM masses, but their values are only, at maximum, $3\%$ larger than their fully rotating counterpart. For the remaining solutions, we observe larger differences between the values of the ADM mass, mainly because of the differences in the angular frequencies, but we see that the fully rotating BSs in the second branch have lower masses than the slowly rotating BSs in their second branch. 

In the right panel, we compare solutions with the same maximal value of the energy-density. Here, slowly rotating BSs with $\text{max}(\rho) \lesssim 10^{-1}$ ($\omega/\mu \gtrsim 0.75$) continue to have slightly more mass than the fully rotating ones, but their agreement is within $3\%$, representing a good approximation. Since slowly rotating BSs can have similar maximum values of the energy-density as the fully rotating BSs -- \textit{cf.} Fig. \ref{Fig:maxrho_omega} -- the difference between both lines is not as pronounced as in the left panel. In fact, by comparing solutions with the same maximal energy-density, the largest deviation in the ADM mass is around $10\%$, except for the last solutions, where the deviation increases to around $30\%$. 
Hence, despite the better agreement than in the left panel, we can confirm again that the approximation holds well enough only for solutions with $\text{max}(\rho) \lesssim 10^{-1}$ ($\omega/\mu \gtrsim 0.75$).

\begin{figure}[ht!]
	\centering
	\includegraphics[width=0.49\linewidth]{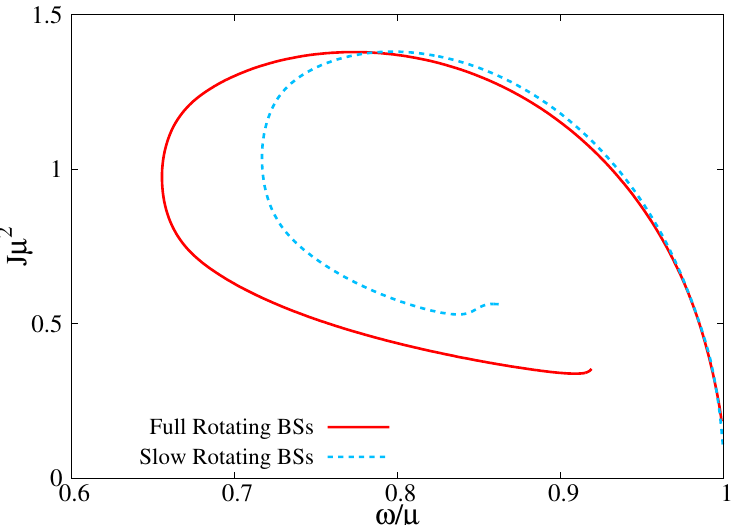}
	\includegraphics[width=0.49\linewidth]{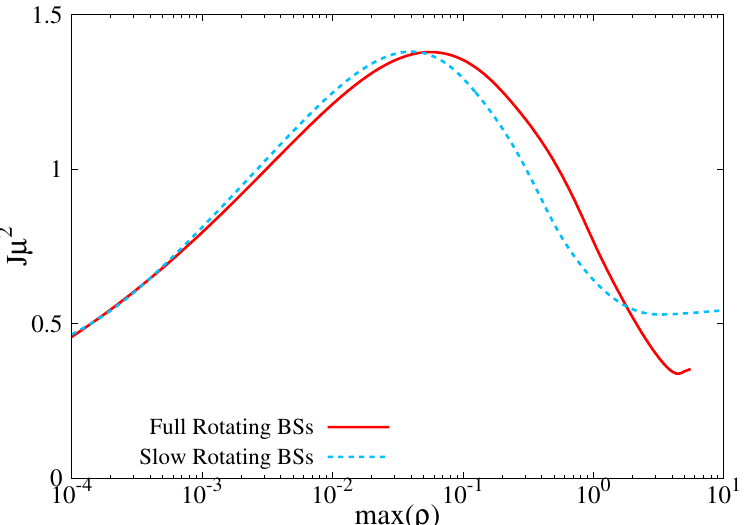}
	\caption{ADM angular momentum, $J\mu^2$, as a function of the angular frequency, $\omega/\mu$, (left panel) and maximum of the energy-density, $\text{max}(\rho)$, (right panel), for fully rotating BSs (red solid line) and slowly rotating BSs (blue dashed line).}
	\label{Fig:AngMom_omega_maxrho}
\end{figure}

We will focus now on the ADM angular momentum. For that, in Fig. \ref{Fig:AngMom_omega_maxrho}, such global charge is shown as a function of both the angular frequency (left panel) and the maximum of the energy-density (right panel) for both families of solutions. 
Here we find very similar results as in the ADM mass. In particular, in both panels, slowly rotating BSs with $\omega/\mu \gtrsim 0.75$ ($\text{max}(\rho) \lesssim 10^{-1}$) present akin (although slightly larger) angular momentum values as the ones for fully rotating BSs. Furthermore, the remaining solutions display larger differences. One of which is the larger minimal angular momentum value of the slowly rotating BSs in the second branch compared to the fully rotating BSs. Such differences leads to the conclusion that the approximation is not good to reproduce the correct values for these solutions, and it is only good enough for BSs with $\omega/\mu \gtrsim 0.75$ ($\text{max}(\rho) \lesssim 10^{-1}$).  

\begin{figure}[ht!]
	\centering
	\includegraphics[width=0.64\linewidth]{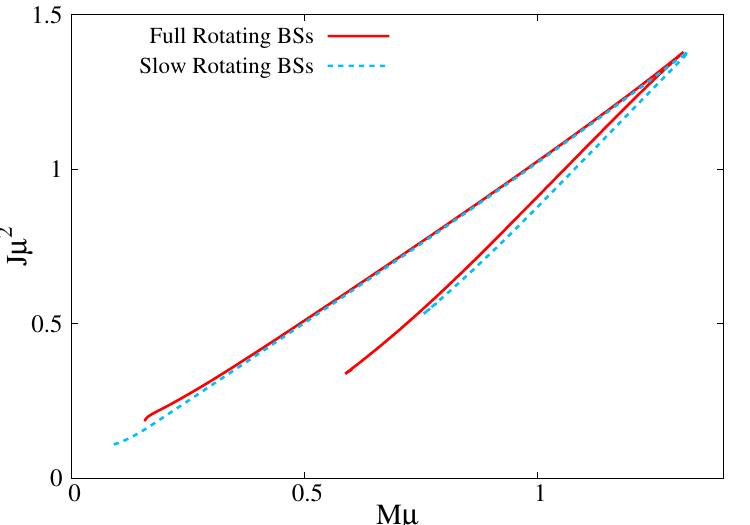}
	\caption{Mass, $M\mu$, \textit{vs} angular momentum, $J\mu^2$ phase space, for fully rotating BSs (red solid line) and slowly rotating BSs (blue dashed line).}
	\label{Fig:Mass_AngMom}
\end{figure}

One can analyze both families of BSs  through their mass \textit{vs.} angular momentum phase space using Fig. \ref{Fig:Mass_AngMom}.
In this plot, we can compare BSs with the same ADM mass. First, we notice the existing correlation between both global charges: as one increases (decreases), the other 
follows suit. Second, we can see a good agreement between both families of solutions, even in the second branch where one would expect larger differences. This implies that, by comparing solutions with the same ADM masses, the approximation introduced in this work can produce slowly rotating BSs whose global quantities are analogous to the fully rotating ones. Third, we find that the bigger difference between both lines is the impossibility to find slowly rotating BSs with low masses and angular momentum as the fully rotating BSs in the second branch. 

\begin{figure}[ht!]
	\centering
	\includegraphics[width=0.49\linewidth]{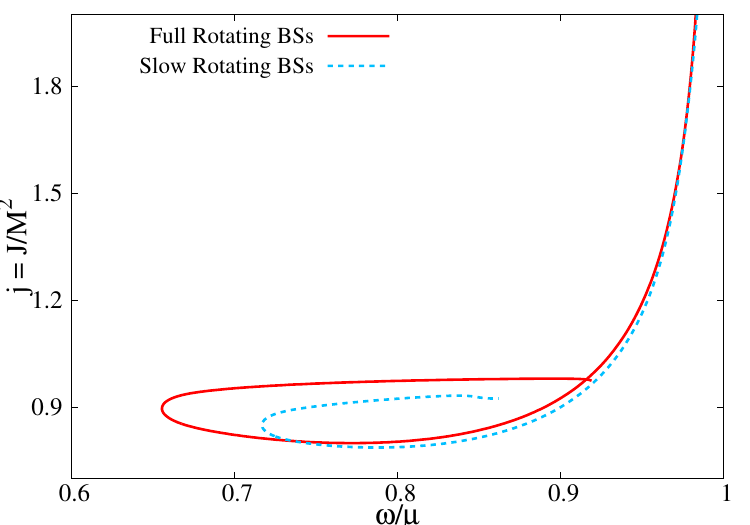}
	\includegraphics[width=0.49\linewidth]{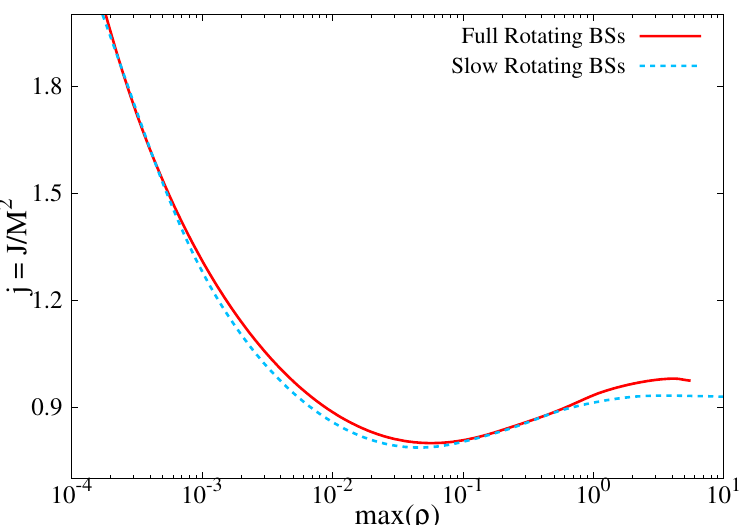}
	\caption{Dimensionless spin, $j = J/M^2$, as a function of the angular frequency, $\omega/\mu$, (left panel) and maximum of the energy-density, $\text{max}(\rho)$, (right panel), for fully rotating BSs (red solid line) and slowly rotating BSs (blue dashed line).}
	\label{Fig:spin_omega_maxrho}
\end{figure}

\begin{figure}[ht!]
	\centering
	\includegraphics[width=0.49\linewidth]{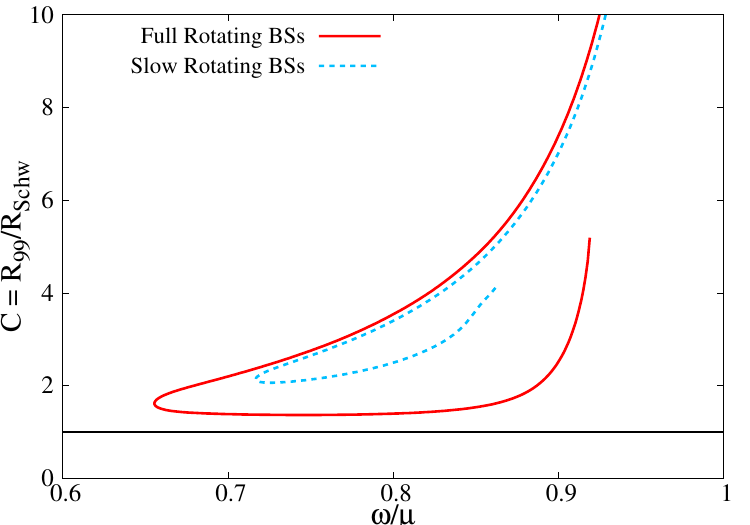}
	\includegraphics[width=0.49\linewidth]{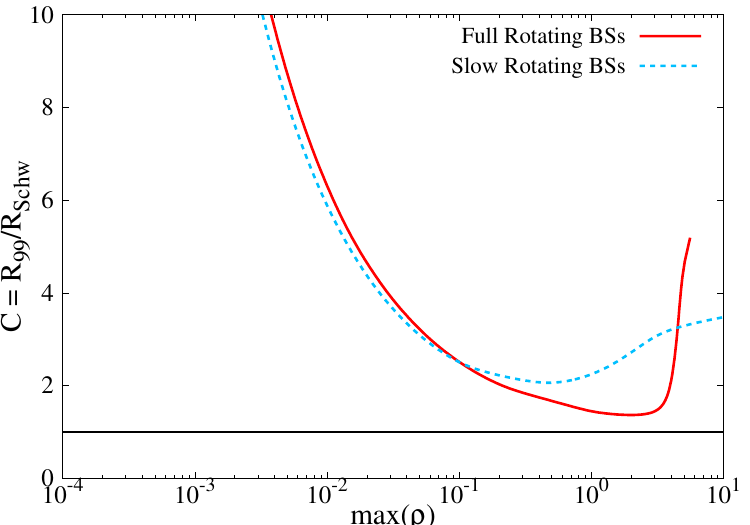}
	\caption{Inverse compactness, $C = R_{99}/R_\text{Schw}$, as a function of the angular frequency, $\omega/\mu$, (left panel) and maximum of the energy-density, $\text{max}(\rho)$, (right panel), for fully rotating BSs (red solid line) and slowly rotating BSs (blue dashed line). The horizontal solid black line at $C = 1$ corresponds to the BH limit.}
	\label{Fig:compactness_omega_maxrho}
\end{figure}

Once the ADM mass and angular momentum are determined one can introduce the dimensionless spin, $j = J/M^2$. This quantity is related closely to the Kerr bound, which for all Kerr black holes (BHs) satisfies the bounds $-1\leq j \leq 1$. 
Fig. \ref{Fig:spin_omega_maxrho} depicts the dimensionless spin as a function of the angular frequency (left panel) and maximal energy-density (right panel). As one moves along the lines away from the Newtonian limit, the solutions become more compact (\textit{cf.} Fig. \ref{Fig:compactness_omega_maxrho}) and the Kerr bound is satisfied.

In Fig. \ref{Fig:spin_omega_maxrho}, we can also compare the dimensionless spin of slowly rotating BSs with fully rotating BSs. 
One can conclude that slowly rotating BSs with $\omega/\mu \gtrsim 0.75$ exhibit spins that are in good agreement with the ones for fully rotating BSs. However, by looking at the right panel, we see a better agreement between the spin values of both families of solutions for almost all possible maximal energy-density values. Indeed, only for solutions with the largest maximal energy-densities we observe larger differences. Nonetheless, such differences do not exceed $5\%$. This shows that by comparing solutions with the same maximal energy-density, the spin of fully rotating BSs are well approximated by the spin of slowly rotating BSs, even in the strong gravity regime.
One can further analyze the compactness of BSs.
It is well-known that BSs do not have a well defined surface  since the scalar field decays exponentially. Nevertheless, we can use the standard procedure~\cite{Amaro-Seoane:2010pks,Herdeiro:2015gia} to defined the ``surface" of a BS.  To do so, we first find the metric coordinate radius, $r_{99}$, that contains $99\%$ of the total mass of the BS, $M_{99}$. 
%
%
We then introduce the
perimetral radius, $R$, related to the metric coordinate radius $r$ by $R := e^{F_1} r$, and compute the corresponding ``surface" $R_{99}$, for which a circumference along the equatorial plane has perimeter $2\pi R_{99}$. 
Finally, we compute the inverse compactness as the ratio of $R_{99}$ and the Schwarzschild radius corresponding to $99\%$ of the total mass, $R_{\text{Schw}} = 2 M_{99}$,
\begin{equation}
	C = \frac{R_{99}}{R_\text{Schw}} = \frac{R_{99}}{2 M_{99}} ~.
\end{equation}
Fig. \ref{Fig:compactness_omega_maxrho} displays the inverse compactness as a function of the angular frequency (left panel) and maximal energy-density (right panel). 
From both panels we can verify that all slowly rotating BSs with $\omega/\mu \gtrsim 0.75$ and $\text{max}(\rho) \lesssim 10^{-1}$ show a compactness similar to fully rotating BSs. Despite that, we still can see that the slowly rotating BSs are a bit more compact than the fully rotating ones. This may be explained by the slightly more massive slowly rotating BSs in this regime, as well as a slightly distinct perimetral radius of the ``surface" of the BS.
The remaining slowly rotating solutions, as we have already seen through the analysis of the previous physical quantities, present large differences compared to the fully rotating ones. 

Another physical quantity that is worth of study is related with the dragging of the spacetime generated by the rotating object -- the frame-dragging effect. To quantify and analyze this, we will measure the angular velocity of a ZAMO on the ``surface" of each BS. 

Let us consider a massive test particle following a geodesic motion around a BS spacetime on the equatorial plane, $\theta = \pi/2$. Its Lagrangian can be express as $2 \mathcal{L} = g_{\mu\nu} \dot{x}^\mu \dot{x}^\nu$, where $x^\mu = \{t,r,\theta,\varphi\}$ are the coordinates of the test particle, and the dot indicates a derivative with respect to an affine parameter, which, for this test particle, is the proper time. 
By Noether's theorem, the two Killing vector fields, $t^a = (\partial_t)^a$ and $\varphi^a = (\partial_\varphi)^a$ present in the spacetime of both families of BSs will lead to two integrals of motion for the test particle. These are the energy, $E$, and angular momentum, $L$,\footnote{The two integrals of motion are normalized by the mass of the test particle that we take as one.} 
\begin{equation}\label{Eq:EnergyAngMom}
	-E = g_{t \mu} \dot{x}^\mu = g_{tt} \dot{t} + g_{t\varphi} \dot{\varphi} ~, \hspace{10pt} L = g_{\varphi \mu} \dot{x}^\mu = g_{t\varphi} \dot{t} + g_{\varphi\varphi} \dot{\varphi} ~.
\end{equation}
The angular velocity associated with the motion of this massive particle can be easily computed by,
\begin{equation}\label{Eq:AngularVelocityCO}
	\Omega = \frac{\dot{\varphi}}{\dot{t}} = - \frac{E g_{t\varphi} + L g_{tt}}{E g_{\varphi\varphi} + L g_{t\varphi}} ~.
\end{equation}
Therefore, for a ZAMO, which, by definition, is an observer whose angular momentum is null, $L = 0$, its angular velocity simplifies to,
\begin{equation}
	\Omega_\text{ZAMO} = -\frac{g_{t\varphi}}{g_{\varphi\varphi}} = \frac{W}{r} ~,
\end{equation}
where, in the last equality, we have used the metric, Eq. \eqref{Eq:AnsatzMetric}, for slowly rotating BSs. Note that, for fully rotating BSs -- \textit{cf.} Eq. \eqref{Eq:AnsatzMetricFull} --, we obtain the same expression, however the function $W$ will be slightly different when compared to the same function for a slowly rotating BS, as one can see in Fig. \ref{Fig:AnsatzFunctions}. Finally, in order to find the angular velocity of a ZAMO at the ``surface" of a BS, we only have to compute the above quantity at $r_{99}$. 

\begin{figure}[ht!]
	\centering
	\includegraphics[width=0.49\linewidth]{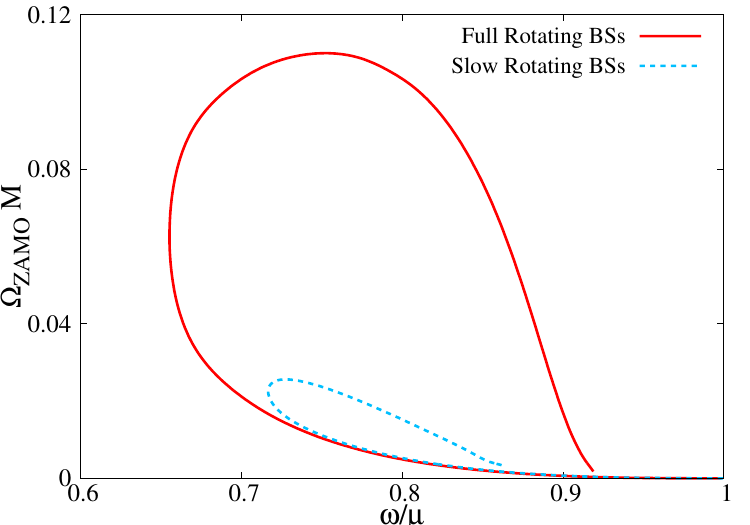}
	\includegraphics[width=0.49\linewidth]{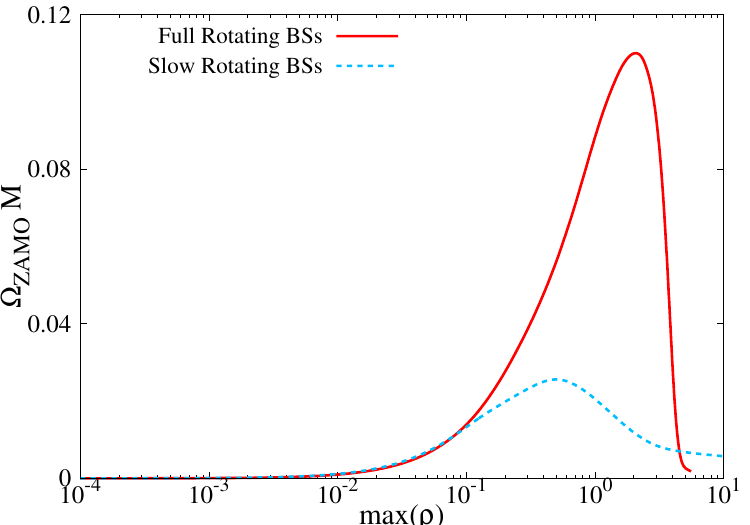}
	\caption{Angular velocity of a ZAMO at the ``surface" of a BS, $\Omega_\text{ZAMO}$, as a function of the angular frequency, $\omega/\mu$, (left panel) and maximum of the energy-density, $\text{max}(\rho)$, (right panel), for fully rotating BSs (red solid line) and slowly rotating BSs (blue dashed line).}
	\label{Fig:Omega99_omega_maxrho}
\end{figure}

In Fig. \ref{Fig:Omega99_omega_maxrho}, we find the angular velocity of a ZAMO at the ``surface" of a BS, as we change either the BS's angular frequency (left panel) or its maximal energy-density (right panel). 
From both panels, we can conclude that any slowly rotating BS with $\omega/\mu \gtrsim 0.75$ and $\text{max}(\rho) \lesssim 10^{-1}$ show an outstanding agreement with the fully rotating BSs. In contrast, all remaining slowly rotating solutions present a very poor agreement. A possible explanation for this result may be due to the strong dependency of $\Omega_\text{ZAMO}$ on $W$. Since as we go along the lines towards the strong gravity regime, the function $W$ start to have larger values, $W \gg 1$, the linear approximation is no longer valid and higher order terms become important. Therefore, the values of the function $W$ for slowly rotating BSs turn to be rather different from the ones for fully rotating BSs, leading to very contrasting angular velocity as the ones shown in Fig. \ref{Fig:Omega99_omega_maxrho}.

Finally, we will analyze the existence of ergo-regions. Their existence is related to some important physical phenomena, for instance, the Penrose process, where orbiting particles can extract rotating energy from the central object~\cite{Penrose:1969pc,Penrose:1971uk}. This process may lead to instabilities in BSs spacetimes, where a full absorbing surface, such as an horizon, is not present~\cite{Cardoso:2007az}.
For fully rotating BSs (both mini-BSs and other models of rotating BSs), and sufficiently compact solutions, ergo-regions develop~\cite{Delgado:2020udb,Kleihaus:2007vk,Herdeiro:2014jaa,Herdeiro:2015gia,Herdeiro:2016gxs,Kunz:2019bhm,Kunz:2019sgn}. These ergo-regions are bounded by an ergo-surface, which, for BSs, have a toroidal topology.

Ergo-regions are defined as the regions of spacetime where the norm of the Killing vector associated with stationarity, $t^a = (\partial_t)^a$, becomes positive, $t^a t_a > 0$. By continuity and due to asymptotically flatness, if an ergo-region exists, then an ergo-surface also exists and it is defined as the place where $t^a t_a = 0$, which, for slowly rotating BSs -- \textit{cf.} Eq. \eqref{Eq:AnsatzMetric} --, can be written as,
\begin{equation}\label{eq:eta1}
    t^a t_a = g_{tt} = -e^{2F_0} + e^{2F_1} W^2 \sin^2 \theta = 0 ~.
\end{equation}
By ignoring the quadratic term of $W$ in \eqref{eq:eta1} one can easily conclude that slowly rotating BSs do not have ergo-regions, since the first term is always negative.
To fix this issue  in slowly rotating BSs we keep track of quadratic terms of $W$ in $g_{tt}$.
This is done following the slowly-rotating approach in a rotating spacetime. For instance, considering 
corrections of first order in $a = J/M$ to the Schwarzschild metric one may conclude that there are no ergoregions. However, Kerr BHs always carry ergo-regions even for small, non-vanishing values of $a$.  

\begin{figure}[ht!]
    \centering
    \includegraphics[width=0.49\linewidth]{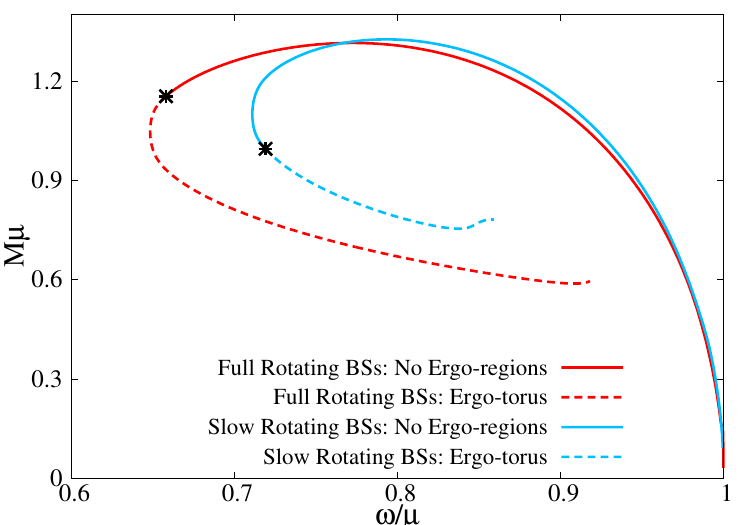}
    \includegraphics[width=0.49\linewidth]{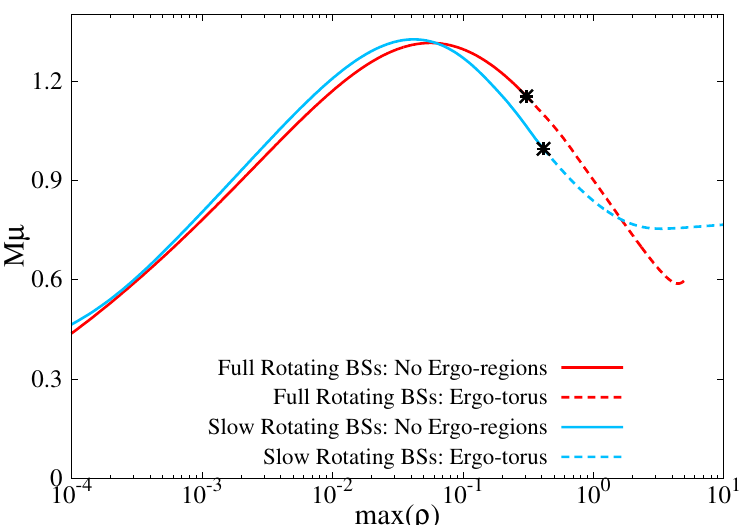}
    \caption{Existence of ergo-regions in an angular frequency, $\omega/\mu$, (left panel) and maximum of the energy-density, $\text{max}(\rho)$, (right panel) \textit{vs.} ADM mass plot. The red (blue) lines correspond to fully (slowly) rotating BSs. Solid (dashed) lines correspond to solutions without (with) ergo-regions. The black stars indicate the first BS that have an ergo-torus.}
    \label{Fig:w_Mass_Ergoregion}
\end{figure}

\begin{figure}[ht!]
    \centering
    \includegraphics[width=0.49\linewidth]{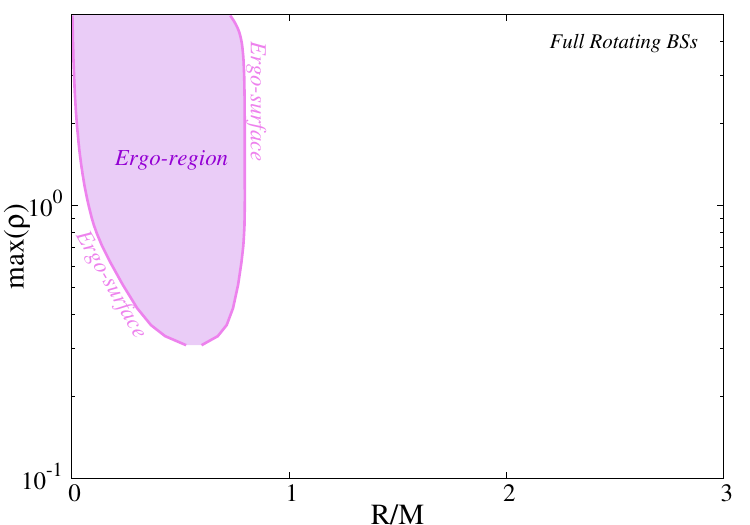}
    \includegraphics[width=0.49\linewidth]{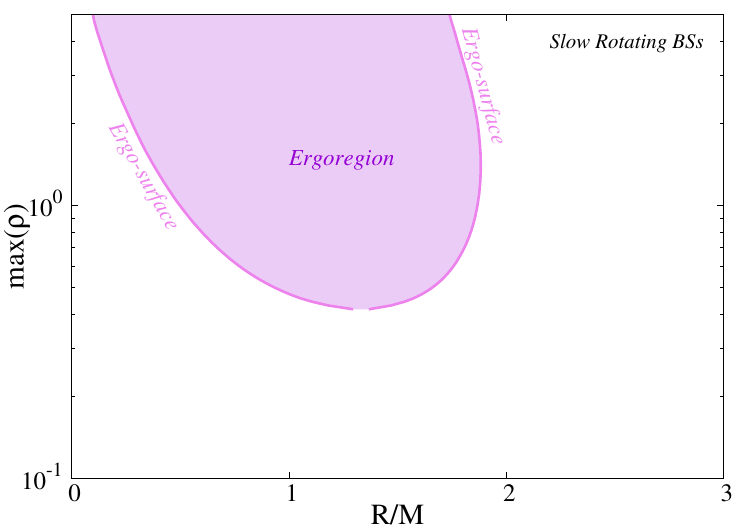}
    \caption{Perimetral radius, $R/M$, of the ergo-surfaces (pink lines) and ergo-regions (violet region) for fully rotating BSs (left panel) and slowly rotating BSs (right panel), computed at the equatorial plane, $\theta = \pi/2$. }
    \label{Fig:StructureErgoregion}
\end{figure}

Fig. \ref{Fig:w_Mass_Ergoregion} shows the existence of ergo-regions for both the fully rotating BSs (red lines) and the slowly rotating BSs (blue lines). 
This is presented in plots of the ADM mass vs.  angular frequency $\omega/\mu$ (left panel) and ADM mass vs. maximum of the energy-density, $\text{max}(\rho)$, (right panel). 
The solid (dashed) lines correspond to solutions without (with) ergo-regions. 
From this figure one can see that ergo-regions appear in slowly rotating BSs in the strong gravity regime where more compact solutions are possible relative to the fully rotating BSs. 
Focusing on the left panel, where one can compare solutions with the same angular frequencies, we conclude that, for both family of BSs, ergo-regions start to develop when we almost reach the solution with the lowest angular frequency value. However, for fully rotating BSs, this happens in the first (top) branch of solutions, whereas, for slowly rotating BSs, it only happens on the second (bottom) branch of solutions. 
If we focus on the right panel, where we can compare solutions with the same maximal energy-density, it is possible to verify that the ergo-regions start to appear for solutions with $\text{max}(\rho) \sim 4 \times 10^{-1}$, for both the fully and slowly rotating BSs.  
Regarding ergo-regions, in Fig. \ref{Fig:StructureErgoregion} we present the ergo-surface and ergo-region, computed at the equatorial plane, $\theta = \pi/2$, for fully rotating BSs (left panel) and slowly rotating BSs (right panel) in a maximal energy-density, $\text{max}(\rho)$, \textit{versus} perimeter-radius $R/M$ plot. In this plot, 
{each constant value of $\text{max}(\rho)$}
corresponds uniquely to one solution. This is useful, since one can see that a solution (both fully or slowly rotating) with $\text{max}(\rho) = 2 \times 10^{-1}$ does not have any ergo-regions, whereas a solution (both fully or slowly rotating) with $\text{max}(\rho) = 1$ have an ergo-region bounded by two ergo-surfaces at two different perimeter radii. 
By comparing both panels, one can conclude that the ergo-regions (and associated ergo-surfaces) of 
fully and slowly rotating BSs appear for different perimetral radii, despite the fact that the ergo-regions start to develop for fully and slowly rotating solutions with similar maximal energy-densities. More concretely, if we fixed the maximal energy-density to $\text{max}(\rho) = 1$, the perimetral radii of the ergo-surfaces of the fully rotating BS are $R/M \sim 0.08$ and $R/M \sim 0.8$, whereas, for a slowly rotating BS with the same maximal energy-density, the perimetral radii are $R/M \sim 0.5$ and $R/M \sim 1.9$. 
Therefore, there is one order of magnitude of difference associated to the perimetral radii of the ergo-surfaces between the fully and slowly rotating BSs. This may be explained because ergo-regions only appear in the strong gravity regime, where the slowly-rotation approximation is not good enough. Nevertheless, it is remarkable that the ergo-regions develop for fully and slowly rotating solutions with very similar maximal energy-densities.

In summary, from the analysis of physical quantities, we can argue that the approximation presented here can produce slowly rotating BSs that yield physical quantities that are in agreement with the ones for fully rotating BSs if the slowly rotating BS has $\omega/\mu \gtrsim 0.75$ and $\text{max}(\rho) \lesssim 10^{-1}$. Outside these values, the difference between the fully and the slowly rotating BSs is noticeable. We can still find similar values for several physical quantities, such as the ADM mass and angular momentum, however, other quantities such as the angular velocity of a ZAMO, can have significantly larger differences.

\subsection{Geodesics around BSs: structure of circular orbits}

We proceed to analyze some phenomenological properties of the slowly rotating BSs by studying the structure of circular orbits (COs) and determine the range of validity of the slowly rotation solution as compared with the fully rotating case. 

The structure of COs can be analyzed by considering the Lagrangian of a test particle moving around a BS in geodesic motion on the equatorial plane,
\begin{equation}
    2\mathcal{L} = g_{\mu\nu} \dot{x}^\mu \dot{x}^\nu = \xi  ~,
\end{equation}
where now $\xi = \{-1,0\}$ for time-like (massive) or light-like (massless) particles. Using Eq. \eqref{Eq:EnergyAngMom}, 
we can write the Lagrangian as~\cite{Delgado:2021jxd,Delgado:2022yvg},
\begin{eqnarray}\label{Eq:LagrangianCOs}
    2\mathcal{L} = - \frac{A(r,E,L)}{B(r)} + g_{rr} \dot{r}^2 = \xi  ~,
\end{eqnarray}
where,
\begin{eqnarray}
    A(r,E,L) \equiv g_{\varphi\varphi} E^2 + 2 g_{t\varphi} E L + g_{tt} L^2 ~ , \hspace{10pt} B(r) \equiv g_{t\varphi}^2 - g_{tt} g_{\varphi\varphi} ~. 
\end{eqnarray}
Note that $B(r) > 0$ throughout all the spacetime generated by the BS, since no horizon exists. This is explained by using the condition $\text{det}(-g) > 0$ together with the positive signature of the $(r,\theta)$-sector of the metric.
We can now introduce the following effective potential,
\begin{equation}
    V_{\text{eff}}(r) \equiv g_{rr} \dot{r}^2 = \xi + \frac{A(r,E,L)}{B(r)} ~.
\end{equation}
Since we want to study COs, the effective potential will be restricted by the following two conditions computed at any COs, $r = r_\text{cir}$,
\begin{eqnarray}
    V_{\text{eff}}(r_\text{cir}) = 0  \hspace{10pt} &\Leftrightarrow& \hspace{10pt} A(r_\text{cir},E,L) = -\xi B(r_\text{cir}) ~, \label{Eq:FirstEqEffPotential}\\
    V_{\text{eff}}'(r_\text{cir}) = 0  \hspace{10pt} &\Leftrightarrow& \hspace{10pt} A'(r_\text{cir},E,L) = -\xi B'(r_\text{cir}) ~, \label{Eq:SecondEqEffPotential}
\end{eqnarray}
where the prime denotes the radial derivative, and we have used the first equation, $V_{\text{eff}}(r_\text{cir}) = 0$, to obtain the final form of the second. After solving the above two equations and finding the existing COs, we can study their radial stability by analysing the sign of the second radial derivative of the effective potential\footnote{Note that we have used Eqs. (\ref{Eq:FirstEqEffPotential}) and (\ref{Eq:SecondEqEffPotential}) to obtain the final form of Eq.(\ref{Eq:StabilityCOs}).},
\begin{equation}\label{Eq:StabilityCOs}
    V_\text{eff}''(r_\text{cir}) = \frac{A''(r_\text{cir},E,L) + \xi B''(r_\text{cir})}{B(r_\text{cir})} ~.
\end{equation}
Then\footnote{{Note that the effective potential used in this work has an opposite sign compared to the standard definition. By studying the epicyclic frequencies using the above potential, one easily sees the presence of a negative sign that yields the inequalities in Eq. \eqref{Eq:StabilityCOs} - \textit{cf.} Eq. (18) in~\cite{Delgado:2022yvg}.}}, 
\begin{equation}
    V_\text{eff}''(r_\text{cir}) > 0 \hspace{5pt} \Leftrightarrow \hspace{5pt} \textit{Unstable CO} \hspace{10pt} ; \hspace{10pt} V_\text{eff}''(r_\text{cir}) < 0 \hspace{5pt} \Leftrightarrow \hspace{5pt} \textit{Stable CO} ~.
\end{equation}
We are now ready to study the structure of COs. Let's start by analyzing the circular motion of light-like particles or light-rings (LRs).
For this class of particles $\xi = 0$, and by introducing the inverse impact parameter, $\sigma = E/L$, Eqs. \eqref{Eq:FirstEqEffPotential} and \eqref{Eq:SecondEqEffPotential} reduce to,
\begin{eqnarray}
    V_{\text{eff}}(r_\text{cir}) = 0  \hspace{10pt} &\Leftrightarrow& \hspace{10pt} g_{\varphi\varphi} \sigma^2 + 2 g_{t\varphi} \sigma + g_{tt}  = 0 ~, \label{Eq:FirstEqLRs} \\
    V_{\text{eff}}'(r_\text{cir}) = 0  \hspace{10pt} &\Leftrightarrow& \hspace{10pt} g_{\varphi\varphi}' \sigma^2 + 2 g_{t\varphi}' \sigma + g_{tt}'  = 0 ~. \label{Eq:SecondEqLRs}
\end{eqnarray}
From Eq. \eqref{Eq:FirstEqLRs} we obtain an expression for the inverse impact parameter entirely written through the metric coefficients,
\begin{equation}
    \sigma = \sigma_\pm = \left. \frac{-g_{t\varphi} \pm \sqrt{B}}{g_{\varphi\varphi}} \right|_\text{LR} ~,
\end{equation}
where $\pm$ denote the two possible rotating senses of the CO. In particular, the $+$ ($-$) corresponds prograde (retrograde) COs. If we combine this result with Eq. \eqref{Eq:SecondEqLRs}, we can obtain the radial coordinate of the LR, $r = r_\text{LR}$.
Finally, the stability of the LR can be easily verified by confirming the sign of Eq. \eqref{Eq:StabilityCOs}, which now reduces to,
\begin{equation}
    V_\text{eff}''(r_\text{LR}) = L^2_\pm \left. \frac{g_{\varphi\varphi}'' \sigma_\pm^2 + 2 g_{t\varphi}'' \sigma_\pm + g_{tt}''}{B} \right|_\text{LR} ~.
\end{equation}
For time-like particles, we can perform a similar study. However, the results are a bit more involved. In order to simplify the results, we will use the angular velocity that was introduced in the previous subsection -- \textit{cf.} Eq. \eqref{Eq:AngularVelocityCO}. 
By using the angular velocity, we can solve Eq. \eqref{Eq:FirstEqEffPotential}, to obtain an expression for the energy and angular momentum written in terms of the metric coefficients and the angular velocity,
\begin{equation}
    E_\pm = \left. - \frac{g_{tt} + g_{t\varphi} \Omega_\pm}{\sqrt{\beta_\pm}} \right|_{r_\text{cir}} ~, \hspace{10pt} L_\pm = \left. \frac{g_{t\varphi} + g_{\varphi\varphi} \Omega_\pm}{\sqrt{\beta_\pm}} \right|_{r_\text{cir}} ~,
\end{equation}
where $\beta_\pm \equiv -(g_{tt} + 2g_{t\varphi}\Omega_\pm + g_{\varphi\varphi} \Omega_\pm^2)|_{r_\text{cir}}$. 
If $\beta_\pm$ yields a positive (zero) [negative] value, then a time-like (light-like) [space-like] CO exists~\cite{Delgado:2021jxd}. 
We can now solve Eq. \eqref{Eq:SecondEqEffPotential} to obtain an expression for the  angular velocity completely composed by the metric coefficients,
\begin{equation}
    \Omega_\pm = \left. \frac{g_{t\varphi}' \pm \sqrt{C}}{g_{\varphi\varphi}'}  \right|_{r_\text{cir}} ~,
\end{equation}
where we have introduced a new function, $C \equiv (g_{t\varphi}')^2 - g_{tt}' g_{\varphi\varphi}'$, that is related to the existence of any kind of COs. A positive (negative) value of $C$ implies the existence (nonexistence) of a CO~\cite{Delgado:2021jxd}. Finally, we can analyze the stability of time-like COs depending on the sign of Eq. \eqref{Eq:StabilityCOs}.
In contrast to the case of the circular motion of light-like particles, where a finite number of LRs exist, the circular motion of time-like particles yields a infinite number of time-like COs (TCOs) which define several regions characterised by the existence and stability of time-like COs,
\begin{itemize}
    \item $C < 0$: Region where it is impossible to find any kind of COs. We dub this region as the \textit{No COs} region.
    \item $C \geq 0~ \wedge~ \beta_\pm < 0$: Region where only space-like COs exist. We dub this region as the \textit{No TCOs} region.
    \item $C \geq 0~ \wedge~ \beta_\pm > 0~ \wedge~ V_\text{eff}'' > 0$: Region where we find unstable time-like COs (UTCOs). We dub this region as the \textit{UTCOs} region.
    \item $C \geq 0~ \wedge~ \beta_\pm > 0~ \wedge~ V_\text{eff}'' < 0$: Region where we find stable time-like COs (STCOs). We dub this region as the \textit{STCOs} region.
\end{itemize}

There are two types of orbits that are at the transition between regions of existence. The first one is the \textit{marginally stable CO} (MSCO). This corresponds to the STCO with the smallest radius that is continually connected to spatial infinity through a sequence of STCOs. For a large set of compact objects, such as Schwarzschild and Kerr BHs, these orbits lay at the transition between the STCOs and UTCOs regions, meaning that the MSCO will correspond to the solution of the following equations that possesses the largest radius,
\begin{equation}\label{Eq:MSCO_ChangeStability}
    V_\text{eff}''(r_\text{MSCO}) = 0 ~~\wedge~~ V_\text{eff}'''(r_\text{MSCO}) < 0 ~.
\end{equation}
However, the MSCO may also be found at the transition between the No COs and STCOs regions. In this case, the MSCO can be obtained by finding the solution of the following equations that possesses the largest radius,
\begin{equation}\label{Eq:MSCO_ChangePossibility}
    C(r_\text{MSCO}) = 0 ~~\wedge ~~ V_\text{eff}''(r_\text{MSCO}) < 0 ~. 
\end{equation}
Although less common than the previous case, it does happen for very hairy BHs with synchronised scalar hair~\cite{Herdeiro:2014goa,Delgado:2023wnj}.

\begin{figure}[ht!]
	\centering
	\includegraphics[width=0.49\linewidth]{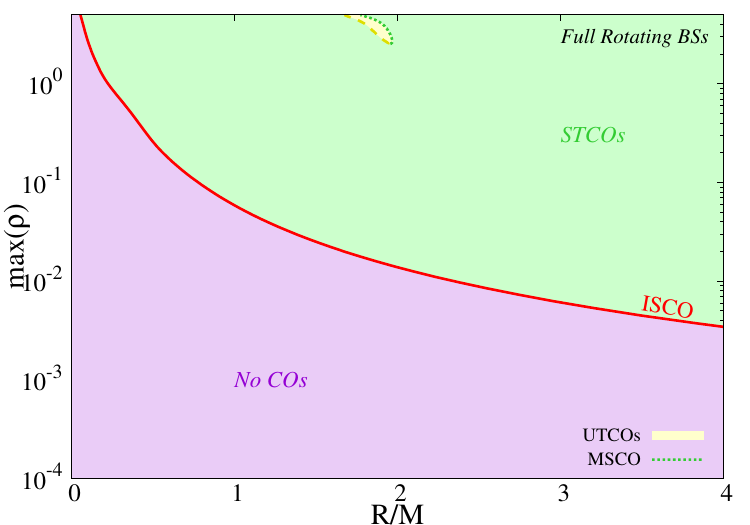}
    \includegraphics[width=0.49\linewidth]{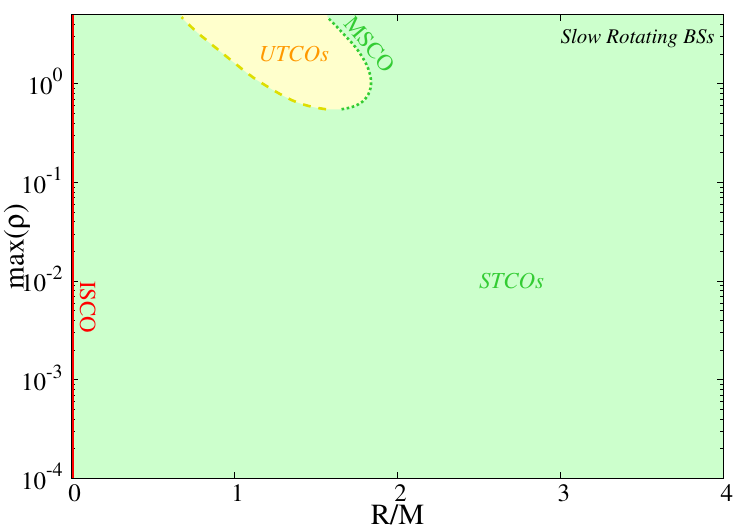}
    \caption{Structure of \textit{prograde} COs around fully rotating (left panel) and slowly rotating BSs (right panel) in a maximal energy-density, $\text{max}(\rho)$, \textit{vs}. perimetral radius, $R/M$. In light green (yellow) [red] we have the region that harbours STCOs (UTCOs) [No TCOs]. The violet region corresponds to the No COs region. The solid red (dotted green) line represents the ISCO (MSCO). Finally, the LRs are presented as a blue dashed line.}
    \label{Fig:StructureCOsPro}
\end{figure}

\begin{figure}[ht!]
	\centering
	\includegraphics[width=0.49\linewidth]{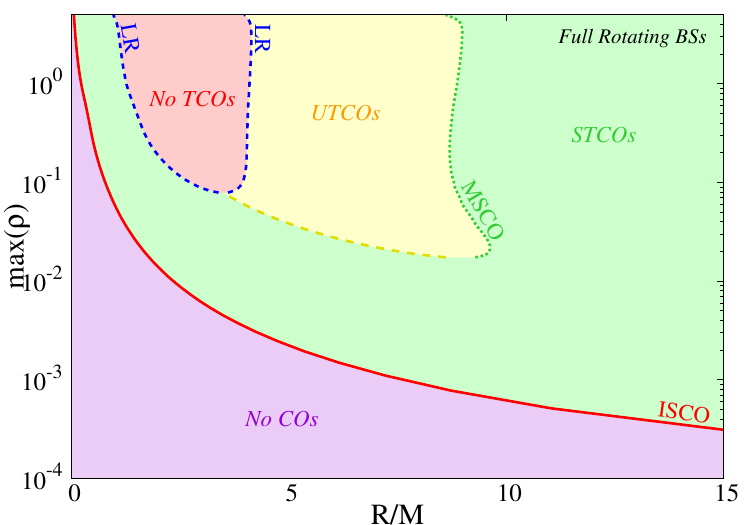}
    \includegraphics[width=0.49\linewidth]{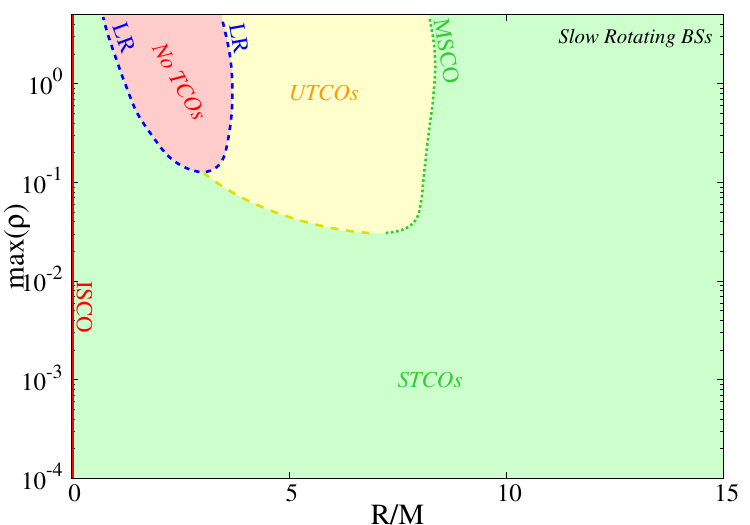}
    \caption{Structure of \textit{retrograde} COs around fully rotating (left panel) and slowly rotating BSs (right panel) in a plot of maximal energy-density, $\text{max}(\rho)$, \textit{vs}. perimetral radius, $R/M$. The color code is the same as in Fig. \ref{Fig:StructureCOsPro}}
    \label{Fig:StructureCOsRetro}
\end{figure}

The second special orbit is the \textit{innermost stable CO} (ISCO). This orbit is similar to the MSCO, however, instead of considering the STCO with the smallest radius that is connected to spatial infinity, we simply consider the STCO that has the smallest possible radius. It can be obtained through the same two ways discussed for the MSCO, Eqs. \eqref{Eq:MSCO_ChangeStability} and \eqref{Eq:MSCO_ChangePossibility}, however, one now searches for the solution with the smallest radius. 
In the Schwarzschild and Kerr BHs case, the ISCO corresponds to the MSCO and it is obtained by solving Eq. \eqref{Eq:MSCO_ChangeStability}. However, for a more exotic object, such as a mini-boson star, the ISCO and the MSCO may be distinct, and, in particular, the former is always found by solving Eq. \eqref{Eq:MSCO_ChangePossibility}~\cite{Delgado:2021jxd}.
We can now focus on the structure of COs for both fully and slowly rotating BSs. Such structure is presented in Fig. \ref{Fig:StructureCOsPro} for prograde orbits and Fig. \ref{Fig:StructureCOsRetro} for retrograde orbits, both in plots of  maximal energy-density, $\text{max}(\rho)$, \textit{versus}, perimetral radius, $R/M$. Similarly to  Fig. \ref{Fig:StructureErgoregion}, these plots are analyzed by drawing horizontal lines, where each line corresponds to a specific BS, and extracting the existing regions and special orbits.
In both figures the color code is the same. In light green (yellow) [red] we have the region that harbours STCOs (UTCOs) [No TCOs]. The violet region corresponds to the No COs region. The solid red (dotted green) line represents the ISCO (MSCO). Finally, the LRs are presented as a blue-dashed line.

Let us start by focusing on the prograde COs -- Fig. \ref{Fig:StructureCOsPro}. On the left panel, we have the structure of COs for the fully rotating BSs, and we can see that for all solutions, there exists always an inner region, containing the origin, where $C < 0$ and no COs can be found. Since this region is connected to a region of STCOs, then, at the beginning of the STCOs region, we find the ISCO. The perimetral radius of the ISCO monotonically decreases as one considers solutions with higher maximal energy-density. For solutions with higher maximal energy-densities, a new region of UTCOs emerges far away from the center of the star. 

On the right panel we performed the same construction of the structure of COs for the slowly rotating BSs. Here one can see a couple of differences. The first is the lack of the inner region where $C < 0$ and no COs exist. This difference is still present for diluted solutions ($\omega/\mu \gtrsim 0.75$ and $\text{max}(\rho) \lesssim 10^{-1}$), where 
no large differences exist between fully and slowly rotating BSs, marking it as the first and only large discrepancy between both family of solutions in this regime. Due to the lack of No COs region, the ISCO is always located at the origin, since we can always have STCOs all the way down to the origin. 
The second difference occurs for solutions with higher values of maximal energy-density. Here we can find a new region of UTCOs and an associated MSCO, similar to the case of the fully rotating BSs. However, the size and location of this new region is quite different when compared to the equivalent UTCOs region of the fully rotating BSs. This may be a consequence of the higher values of the $W$ function.
Regarding retrograde orbits, they are described in Fig. \ref{Fig:StructureCOsRetro}. For fully rotating BSs (left panel), one can distinguish the inner region of No COs and ISCO. Indeed, this region and the orbits are independent of the rotating sense of the CO, since when $C = 0$, the angular velocity is the same for both rotating directions, $\Omega_+ = \Omega_- = -g_{t\varphi}'/g_{\varphi\varphi}'$. The larger difference, as compared to the prograde case, is the larger region of UTCOs, and the existence of a pair of LRs with opposite stabilities~\cite{Cunha:2017qtt}, which always brings with it a region of No TCOs~\cite{Delgado:2021jxd,Delgado:2022yvg}.

For slowly rotating BSs (right panel), it is possible to verify the nonexistence of a No COs region, similar to what we already discuss in the prograde case, marking it as the biggest difference between both families of solutions. Nonetheless, the slowly rotating BSs still show a similar structure regarding the existence of LRs and the regions of UTCOs and No TCOs. However, the sizes and radii of these regions are different from the ones for fully rotating BSs, mainly due the fact that these regions only appear for solutions in the strong gravity regime ($\omega/\mu \lesssim 0.75$ and $\text{max}(\rho) \gtrsim 10^{-1}$). In the end, without taking into account the nonexistence of the region of No COs, the slowly rotating BSs can mimic qualitatively the structure of COs for the fully rotating BSs.

\section{Conclusion}
\label{sec:final}

In the present work we have computed slowly rotating BSs and studied their similarities with fully rotating BSs, both families being
horizonless, regular everywhere, stationary and asymptotically flat solutions of the EKG theory, Eq.\eqref{Eq:Action}.
In the slow rotation approximation, however, we only consider linear terms in
the metric function $W$, which contains the spacetime information regarding the rotation of a given solution. 

We started by showing, in a similar way to the work by Ref.\cite{Kobayashi:1994qi}, that a fully consistent slowly rotating solution is impossible to find 
{assuming axial symmetry}. This is true because the equation $E_r^\theta = 0$ implies that the angular part of the scalar field, which is described by the spherical harmonics, Eq. \eqref{Eq:SphericalHarmonicsEquation}, must be a constant. This is only possible when $l=m=0$. In turn, such result, together with asymptotic flatness, entails that the $W$ function is trivial. Hence, no slowly rotating BSs exist. However, it is possible to devise a strategy that opens the possibility to consider a different solution for the angular part of the scalar field. In particular, the spherical harmonic with $l=m=1$. By considering this spherical harmonic and performing an average of the system of equations over the solid angle, Eq. \eqref{Eq:WeightedAverage}, the full system is completely consistent and can yield non-trivial solutions for the $W$ function. 

Furthermore, we have studied some of the physical quantities and the structure of COs around the slowly rotating BSs and compared their results with the fully rotating case. Regarding the physical quantities, we verified that slowly rotating solutions with $\omega/\mu \gtrsim 0.75$ and $\text{max}(\rho) \lesssim 10^{-1}$ yield very similar results as their fully rotating counterpart. This was expected since for such solutions $W \ll 1$ and aforementioned average is sufficiently good. 
The largest differences here resides on the range of values of the angular frequency, $\omega/\mu$ and the angular velocity of a ZAMO at the ``surface" of a BS. In the former, the slowly rotating BSs can have a minimal frequency as the fully rotating BSs, but this minimal frequency is larger than the minimal frequency of the fully rotating BSs. In the latter, slowly rotating solutions with $\text{max}(\rho) \gtrsim 10^{-1}$ show a considerable discrepancy compared to their fully rotating configurations that is not seen on the remaining physical quantities. 

Regarding the structure of COs, we saw that slowly rotating BSs can display a similar structure as the fully rotating BSs, even for solutions in the strong gravity regime, where the approximation would be compromised. However, such statement is only true if we ignore the lack of an inner region of No COs on the slowly rotating BSs. Such inner region is highly characteristic of fully rotating BSs, since they are one of the few family of exotic compact objects that possess it. The lack of this inner region still remains when considering diluted solutions ($\omega/\mu \gtrsim 0.75$ and $\text{max}(\rho) \lesssim 10^{-1}$), thus marking the only large discrepancy between both families of BSs in such regime of solutions.

In summary, the strategy presented in this work to compute slowly rotating BSs allows one to simplify the system of partial differential equations into a system of ordinary differential equations, where solutions are easier to obtain and yield good approximations to the fully rotating scenario, notably, in 
most of the physical quantities analyzed, and in the regime where $\omega/\mu \gtrsim 0.75$ and $\text{max}(\rho) \lesssim 10^{-1}$. 
Due to the similarities between both families of rotating BSs and the 
simplifications 
that entails the study of static and slowly rotating objects, 
the strategy followed in this work may be applied in the future to analyze the influence of rotation in other systems, such as the analysis of stability, time evolution and gravitational waves produced by the interaction of exotic compact objects.


\section{Acknowledgments}

This work was partially supported by 
DGAPA-UNAM through grants IN110523, IN105223, Conacyt grant 140630, by UNAM Posdoctoral Program (POSDOC),
by the CONACyT Network Project No. 376127 ``Sombras, lentes y ondas gravitatorias generadas por objetos compactos astrof\'\i sicos" and No. 304001 ``Estudio de campos escalares con aplicaciones en cosmolog\'ia y astrof\'isica"
and by the programme HORIZON-MSCA2021-SE-01 Grant No. NewFun-FiCO101086251.


\bibliography{bibliography}

\begin{thebibliography}{10}

\bibitem{Kaup:1968zz}
D.~J. Kaup, ``{Klein-Gordon Geon},'' {\em Phys. Rev.}, vol.~172,
  pp.~1331--1342, 1968.

\bibitem{Ruffini:1969qy}
R.~Ruffini and S.~Bonazzola, ``{Systems of selfgravitating particles in general
  relativity and the concept of an equation of state},'' {\em Phys. Rev.},
  vol.~187, pp.~1767--1783, 1969.

\bibitem{Colpi}
M.~Colpi, S.~L. Shapiro, and I.~Wasserman, ``{Gravitational equilibria of
  self-interacting scalar fields},'' {\em Phys. Rev. D}, vol.~57,
  pp.~2485--2488, 1986.

\bibitem{Jetzer:1991jr}
P.~Jetzer, ``{Boson stars},'' {\em Phys. Rept.}, vol.~220, pp.~163--227, 1992.

\bibitem{Schunck:2003kk}
F.~E. Schunck and E.~W. Mielke, ``{General relativistic boson stars},'' {\em
  Class. Quant. Grav.}, vol.~20, pp.~R301--R356, 2003.

\bibitem{Liebling:2012fv}
S.~L. Liebling and C.~Palenzuela, ``{Dynamical boson stars},'' {\em Living Rev.
  Rel.}, vol.~26, no.~1, p.~1, 2023.

\bibitem{Kobayashi:1994qi}
Y.~Kobayashi, M.~Kasai, and T.~Futamase, ``{Does a boson star rotate?},'' {\em
  Phys. Rev. D}, vol.~50, pp.~7721--7724, 1994.

\bibitem{Yoshida1997a}
S.~Yoshida and Y.~Eriguchi, ``{New static axisymmetric and nonvacuum solutions
  in general relativity: Equilibrium solutions of boson stars},'' {\em Phys.
  Rev. D}, vol.~55, pp.~2000--, 1997.

\bibitem{Schunck:1996he}
F.~E. Schunck and E.~W. Mielke, ``{Rotating boson star as an effective mass
  torus in general relativity},'' {\em Phys. Lett. A}, vol.~249, pp.~389--394,
  1998.

\bibitem{Yoshida:1997qf}
S.~Yoshida and Y.~Eriguchi, ``{Rotating boson stars in general relativity},''
  {\em Phys. Rev. D}, vol.~56, pp.~762--771, 1997.

\bibitem{Grandclement:2014msa}
P.~Grandclement, C.~Som\'e, and E.~Gourgoulhon, ``{Models of rotating boson
  stars and geodesics around them: new type of orbits},'' {\em Phys. Rev. D},
  vol.~90, no.~2, p.~024068, 2014.

\bibitem{Ontanon:2021hbg}
S.~Ontanon and M.~Alcubierre, ``{Rotating boson stars using finite differences
  and global Newton methods},'' {\em Class. Quant. Grav.}, vol.~38, no.~15,
  p.~154003, 2021.

\bibitem{Matos:1998vk}
T.~Matos and F.~S. Guzman, ``{Scalar fields as dark matter in spiral
  galaxies},'' {\em Class. Quant. Grav.}, vol.~17, pp.~L9--L16, 2000.

\bibitem{Matos:1999et}
T.~Matos, F.~S. Guzman, and L.~A. Urena-Lopez, ``{Scalar field as dark matter
  in the universe},'' {\em Class. Quant. Grav.}, vol.~17, pp.~1707--1712, 2000.

\bibitem{Matos:2000ss}
T.~Matos and L.~A. Urena-Lopez, ``{A Further analysis of a cosmological model
  of quintessence and scalar dark matter},'' {\em Phys. Rev. D}, vol.~63,
  p.~063506, 2001.

\bibitem{Hui:2016ltb}
L.~Hui, J.~P. Ostriker, S.~Tremaine, and E.~Witten, ``{Ultralight scalars as
  cosmological dark matter},'' {\em Phys. Rev. D}, vol.~95, no.~4, p.~043541,
  2017.

\bibitem{Mielke:2000mh}
E.~W. Mielke and F.~E. Schunck, ``{Boson stars: Alternatives to primordial
  black holes?},'' {\em Nucl. Phys. B}, vol.~564, pp.~185--203, 2000.

\bibitem{Klaer:2017ond}
V.~B.~. Klaer and G.~D. Moore, ``{The dark-matter axion mass},'' {\em JCAP},
  vol.~11, p.~049, 2017.

\bibitem{Schunck:1997dn}
F.~E. Schunck and A.~R. Liddle, ``{The Gravitational redshift of boson
  stars},'' {\em Phys. Lett. B}, vol.~404, pp.~25--32, 1997.

\bibitem{Torres:2000dw}
D.~F. Torres, S.~Capozziello, and G.~Lambiase, ``{A Supermassive scalar star at
  the galactic center?},'' {\em Phys. Rev. D}, vol.~62, p.~104012, 2000.

\bibitem{Lemos:2008cv}
J.~P.~S. Lemos and O.~B. Zaslavskii, ``{Black hole mimickers: Regular versus
  singular behavior},'' {\em Phys. Rev. D}, vol.~78, p.~024040, 2008.

\bibitem{Mazur:2001fv}
P.~O. Mazur and E.~Mottola, ``{Gravitational Condensate Stars: An Alternative
  to Black Holes},'' {\em Universe}, vol.~9, no.~2, p.~88, 2023.

\bibitem{Guzman:2005bs}
F.~S. Guzman, ``{Accretion disc onto boson stars: A Way to supplant black holes
  candidates},'' {\em Phys. Rev. D}, vol.~73, p.~021501, 2006.

\bibitem{Vincent:2015xta}
F.~H. Vincent, Z.~Meliani, P.~Grandclement, E.~Gourgoulhon, and O.~Straub,
  ``{Imaging a boson star at the Galactic center},'' {\em Class. Quant. Grav.},
  vol.~33, no.~10, p.~105015, 2016.

\bibitem{Meliani:2015zta}
Z.~Meliani, F.~H. Vincent, P.~Grandcl\'ement, E.~Gourgoulhon,
  R.~Monceau-Baroux, and O.~Straub, ``{Circular geodesics and thick tori around
  rotating boson stars},'' {\em Class. Quant. Grav.}, vol.~32, no.~23,
  p.~235022, 2015.

\bibitem{Sanchis-Gual:2018oui}
N.~Sanchis-Gual, C.~Herdeiro, J.~A. Font, E.~Radu, and F.~Di~Giovanni,
  ``{Head-on collisions and orbital mergers of Proca stars},'' {\em Phys. Rev.
  D}, vol.~99, no.~2, p.~024017, 2019.

\bibitem{Grandclement:2016eng}
P.~Grandcl\'ement, ``{Light rings and light points of boson stars},'' {\em
  Phys. Rev. D}, vol.~95, no.~8, p.~084011, 2017.

\bibitem{Zhang:2021xhp}
Y.-P. Zhang, Y.-B. Zeng, Y.-Q. Wang, S.-W. Wei, and Y.-X. Liu, ``{Motion of
  test particle in rotating boson star},'' {\em Phys. Rev. D}, vol.~105, no.~4,
  p.~044021, 2022.

\bibitem{Olivares:2018abq}
H.~Olivares, Z.~Younsi, C.~M. Fromm, M.~De~Laurentis, O.~Porth, Y.~Mizuno,
  H.~Falcke, M.~Kramer, and L.~Rezzolla, ``{How to tell an accreting boson star
  from a black hole},'' {\em Mon. Not. Roy. Astron. Soc.}, vol.~497, no.~1,
  pp.~521--535, 2020.

\bibitem{Wald}
R.~M. Wald, {\em General Relativity}.
\newblock University of Chicago Press, 1984.

\bibitem{Herdeiro:2016gxs}
C.~A.~R. Herdeiro, E.~Radu, and H.~F. R\'unarsson, ``{Spinning boson stars and
  Kerr black holes with scalar hair: the effect of self-interactions},'' {\em
  Int. J. Mod. Phys. D}, vol.~25, no.~09, p.~1641014, 2016.

\bibitem{Delgado:2020udb}
J.~F.~M. Delgado, C.~A.~R. Herdeiro, and E.~Radu, ``{Rotating Axion Boson
  Stars},'' {\em JCAP}, vol.~06, p.~037, 2020.

\bibitem{Delgado:2019prc}
J.~F.~M. Delgado, C.~A.~R. Herdeiro, and E.~Radu, ``{Kerr black holes with
  synchronised scalar hair and higher azimuthal harmonic index},'' {\em Phys.
  Lett. B}, vol.~792, pp.~436--444, 2019.

\bibitem{Delgado:2016jxq}
J.~F.~M. Delgado, C.~A.~R. Herdeiro, E.~Radu, and H.~Runarsson,
  ``{Kerr\textendash{}Newman black holes with scalar hair},'' {\em Phys. Lett.
  B}, vol.~761, pp.~234--241, 2016.

\bibitem{Herdeiro:2021jgc}
C.~Herdeiro, I.~Perapechka, E.~Radu, and Y.~Shnir, ``{Spinning gauged boson and
  Dirac stars: A comparative study},'' {\em Phys. Lett. B}, vol.~824,
  p.~136811, 2022.

\bibitem{Li:2020ffy}
H.-B. Li, Y.-B. Zeng, Y.~Song, and Y.-Q. Wang, ``{Self-interacting multistate
  boson stars},'' {\em JHEP}, vol.~04, p.~042, 2021.

\bibitem{Zeng:2023hvq}
Y.-B. Zeng, S.-X. Sun, S.-Y. Cui, Y.-P. Zhang, and Y.-Q. Wang, ``{Rotating
  multistate axion boson stars},'' 9 2023.

\bibitem{Adam:2022nlq}
C.~Adam, J.~Castelo, A.~Garc\'\i{}a Mart\'\i{}n-Caro, M.~Huidobro,
  R.~V\'azquez, and A.~Wereszczynski, ``{Universal relations for rotating boson
  stars},'' {\em Phys. Rev. D}, vol.~106, no.~12, p.~123022, 2022.

\bibitem{Hartle}
J.~Hartle, ``Slowly rotating relativistic stars. {I}. equations of structure,''
  {\em The Astrophysical Journal}, vol.~150, p.~1005, 1967.

\bibitem{Herdeiro:2014goa}
C.~A.~R. Herdeiro and E.~Radu, ``{Kerr black holes with scalar hair},'' {\em
  Phys. Rev. Lett.}, vol.~112, p.~221101, 2014.

\bibitem{Herdeiro:2015gia}
C.~Herdeiro and E.~Radu, ``{Construction and physical properties of Kerr black
  holes with scalar hair},'' {\em Class. Quant. Grav.}, vol.~32, no.~14,
  p.~144001, 2015.

\bibitem{Delgado:2020hwr}
J.~F.~M. Delgado, C.~A.~R. Herdeiro, and E.~Radu, ``{Kerr black holes with
  synchronized axionic hair},'' {\em Phys. Rev. D}, vol.~103, no.~10,
  p.~104029, 2021.

\bibitem{Alcubierre:2021psa}
M.~Alcubierre, J.~Barranco, A.~Bernal, J.~C. Degollado, A.~Diez-Tejedor,
  V.~Jaramillo, M.~Megevand, D.~N\'u\~nez, and O.~Sarbach, ``{Extreme
  \ensuremath{\ell}-boson stars},'' {\em Class. Quant. Grav.}, vol.~39, no.~9,
  p.~094001, 2022.

\bibitem{Poisson_2004}
E.~Poisson, {\em A Relativist’s Toolkit: The Mathematics of Black-Hole
  Mechanics}.
\newblock Cambridge University Press, 2004.

\bibitem{Amaro-Seoane:2010pks}
P.~Amaro-Seoane, J.~Barranco, A.~Bernal, and L.~Rezzolla, ``{Constraining
  scalar fields with stellar kinematics and collisional dark matter},'' {\em
  JCAP}, vol.~11, p.~002, 2010.

\bibitem{Penrose:1969pc}
R.~Penrose, ``{Gravitational collapse: The role of general relativity},'' {\em
  Riv. Nuovo Cim.}, vol.~1, pp.~252--276, 1969.

\bibitem{Penrose:1971uk}
R.~Penrose and R.~M. Floyd, ``{Extraction of rotational energy from a black
  hole},'' {\em Nature}, vol.~229, pp.~177--179, 1971.

\bibitem{Cardoso:2007az}
V.~Cardoso, P.~Pani, M.~Cadoni, and M.~Cavaglia, ``{Ergoregion instability of
  ultracompact astrophysical objects},'' {\em Phys. Rev. D}, vol.~77,
  p.~124044, 2008.

\bibitem{Kleihaus:2007vk}
B.~Kleihaus, J.~Kunz, M.~List, and I.~Schaffer, ``{Rotating Boson Stars and
  Q-Balls. II. Negative Parity and Ergoregions},'' {\em Phys. Rev. D}, vol.~77,
  p.~064025, 2008.

\bibitem{Herdeiro:2014jaa}
C.~Herdeiro and E.~Radu, ``{Ergosurfaces for Kerr black holes with scalar
  hair},'' {\em Phys. Rev. D}, vol.~89, no.~12, p.~124018, 2014.

\bibitem{Kunz:2019bhm}
J.~Kunz, I.~Perapechka, and Y.~Shnir, ``{Kerr black holes with parity-odd
  scalar hair},'' {\em Phys. Rev. D}, vol.~100, no.~6, p.~064032, 2019.

\bibitem{Kunz:2019sgn}
J.~Kunz, I.~Perapechka, and Y.~Shnir, ``{Kerr black holes with synchronised
  scalar hair and boson stars in the Einstein-Friedberg-Lee-Sirlin model},''
  {\em JHEP}, vol.~07, p.~109, 2019.

\bibitem{Delgado:2021jxd}
J.~F.~M. Delgado, C.~A.~R. Herdeiro, and E.~Radu, ``{Equatorial timelike
  circular orbits around generic ultracompact objects},'' {\em Phys. Rev. D},
  vol.~105, no.~6, p.~064026, 2022.

\bibitem{Delgado:2022yvg}
J.~F.~M. Delgado, ``{Epicyclic frequencies for a generic ultracompact
  object},'' {\em Phys. Rev. D}, vol.~106, no.~6, p.~064054, 2022.

\bibitem{Delgado:2023wnj}
J.~F.~M. Delgado, C.~A.~R. Herdeiro, and E.~Radu, ``{EMRIs around j = 1 black
  holes with synchronised hair},'' {\em JCAP}, vol.~10, p.~029, 2023.

\bibitem{Cunha:2017qtt}
P.~V.~P. Cunha, E.~Berti, and C.~A.~R. Herdeiro, ``{Light-Ring Stability for
  Ultracompact Objects},'' {\em Phys. Rev. Lett.}, vol.~119, no.~25, p.~251102,
  2017.

\end{thebibliography}
\bibliographystyle{ieeetr}

\end{document}